\documentclass{gGAF2e}
\usepackage[colorlinks,allcolors=blue]{hyperref}
\usepackage{mathrsfs}
\usepackage{gensymb}
\DeclareMathAlphabet{\mathscrbf}{OMS}{mdugm}{b}{n}
\graphicspath{{./fig/}{./png/}}
\begin{document}
\jvol{00} \jnum{00} \jyear{2018} 

\markboth{\rm{P.J.\ K\"APYL\"A et al.}}{\rm GEOPHYSICAL AND ASTROPHYSICAL FLUID DYNAMICS}

\renewcommand{\Omega}{{\varOmega}}
\renewcommand{\Gamma}{{\varGamma}}
\newcommand{\DR}{{\mathrm D}}



\newcommand{\aaa}{{\bm a}}
\newcommand{\maaa}{\overline{\bm a}}
\newcommand{\bbb}{{\bm b}}
\newcommand{\mbbb}{\overline{\bm b}}
\newcommand{\eee}{{\bm e}}
\newcommand{\meee}{\overline{\bm e}}
\newcommand{\fff}{{\bm f}}
\newcommand{\mfff}{\overline{\bm f}}
\newcommand{\jjj}{{\bm j}}
\newcommand{\mjjj}{\overline{\bm j}}
\newcommand{\kkk}{{\bm k}}
\newcommand{\mkkk}{\overline{\bm k}}
\newcommand{\rrr}{{\bm r}}
\newcommand{\mrrr}{\overline{\bm r}}
\newcommand{\uuu}{{\bm u}}
\newcommand{\muuu}{\overline{\bm u}}
\newcommand{\xxx}{{\bm x}}
\newcommand{\aaaT}{{\bm a}^{\rm T}}
\newcommand{\bbbT}{{\bm b}^{\rm T}}
\newcommand{\jjjT}{{\bm j}^{\rm T}}
\newcommand{\ooo}{{\bm \omega}}
\newcommand{\AAA}{{\bm A}}
\newcommand{\mAAA}{\overline{\bm A}}
\newcommand{\AAAT}{{\bm A}^{\rm T}}
\newcommand{\BBB}{{\bm B}}
\newcommand{\mBBB}{\overline{\bm B}}
\newcommand{\BBBT}{{\bm B}^{\rm T}}
\newcommand{\mBBBT}{\overline{\bm B}^{\rm T}}
\newcommand{\FFF}{{\bm F}}
\newcommand{\mFFF}{\overline{\bm F}}
\newcommand{\FFFT}{{\bm F}^{\rm T}}
\newcommand{\JJJ}{{\bm J}}
\newcommand{\mJJJ}{\overline{\bm J}}
\newcommand{\JJJT}{{\bm J}^{\rm T}}
\newcommand{\UUU}{{\bm U}}
\newcommand{\mUUU}{\overline{\bm U}}
\newcommand{\UUUT}{{\bm U}^{\rm T}}
\newcommand{\msFFF}{\mathscrbf{F}}
\newcommand{\mfi}[2]{\overline{#1}_{#2}}
\newcommand{\mfvi}[2]{\overline{\bm #1}_{#2}}
\newcommand{\mUUi}{\overline{U}_i}
\newcommand{\mUUj}{\overline{U}_l}
\newcommand{\mUUk}{\overline{U}_k}
\newcommand{\mBBi}{\overline{B}_i}
\newcommand{\mBBj}{\overline{B}_j}
\newcommand{\mBBk}{\overline{B}_k}
\newcommand{\mJJi}{\overline{J}_i}
\newcommand{\mJJj}{\overline{J}_j}
\newcommand{\mJJk}{\overline{J}_k}
\newcommand{\mEMF}{\overline{\bm{\mathcal{E}}}}
\newcommand{\EMFi}{\overline{\mathcal{E}}_i}
\newcommand{\SSt}{\bm{\mathsf{S}}}
\newcommand{\SStij}{\mathsf{S}_{ij}}
\newcommand{\alij}{\alpha_{ij}}
\newcommand{\etij}{\eta_{ij}}
\newcommand{\beijk}{\beta_{ijk}}
\newcommand{\Eq}[1]{Eq.~(\ref{#1})}
\newcommand{\Equ}[1]{Equation~(\ref{#1})}
\newcommand{\equ}[1]{equation~(\ref{#1})}
\newcommand{\eqs}[2]{equations~(\ref{#1}) and (\ref{#2})}
\newcommand{\Eqs}[2]{Equations~(\ref{#1}) and (\ref{#2})} 
\newcommand{\Equs}[2]{Equations~(\ref{#1}) to (\ref{#2})}
\newcommand{\Eqsa}[2]{Eqs.~(\ref{#1}) and (\ref{#2})} 
\newcommand{\Equsa}[2]{Eqs.~(\ref{#1}) to (\ref{#2})}
\newcommand{\EQ}{\begin{equation}}
\newcommand{\EN}{\end{equation}}
\newcommand{\EQA}{\begin{eqnarray}}
\newcommand{\ENA}{\end{eqnarray}}
\newcommand{\brac}[1]{\langle #1 \rangle}
\newcommand{\pd}{\upartial}
\newcommand{\pdz}{\upartial_z}
\newcommand{\DIV}{\vec{\nabla} \cdot }
\newcommand{\CURL}{\vec{\nabla} \times }
\newcommand{\cross}[2]{\boldsymbol{#1} \times \boldsymbol{#2}}
\newcommand{\crossm}[2]{\brac{\boldsymbol{#1}} \times \brac{\boldsymbol{#2}}}
\newcommand{\ve}[1]{\boldsymbol{#1}}
\newcommand{\mean}[1]{\overline{#1}}
\newcommand{\meanv}[1]{\overline{\bm #1}}
\newcommand{\cP}{c_{\rm P}}
\newcommand{\cV}{c_{\rm V}}
\newcommand{\cs}{c_{\rm s}}
\newcommand{\cst}{c_{\rm s}^2}
\newcommand{\csk}{c_{\rm s}^3}
\newcommand{\nut}{\nu_{\rm t}}
\newcommand{\nutz}{\nu_{\rm t0}}
\newcommand{\tnut}{\tilde{\nu}_{\rm t}}
\newcommand{\etat}{\eta_{\rm t}}
\newcommand{\etatz}{\eta_{\rm t0}}
\newcommand{\etaT}{\eta_{\rm T}}
\newcommand{\tetat}{\tilde{\eta}_{\rm t}}
\newcommand{\urms}{u_{\rm rms}}
\newcommand{\Urms}{U_{\rm rms}}
\newcommand{\brms}{B_{\rm rms}}
\newcommand{\orms}{\omega_{\rm rms}}
\newcommand{\Beq}{B_{\rm eq}}
\newcommand{\Ma}{{\rm Ma}}
\newcommand{\sigmat}{\sigma_{\rm t}}
\newcommand{\eu}{\hat{\bm e}}
\newcommand{\xu}{\hat{\bm x}}
\newcommand{\yu}{\hat{\bm y}}
\newcommand{\zu}{\hat{\bm z}}
\newcommand{\ru}{\hat{\bm r}}
\newcommand{\Ou}{\hat{\bm \Omega}}
\newcommand{\kB}{k_B}
\newcommand{\ku}{k_u}
\newcommand{\kf}{k_{\rm f}}
\newcommand{\tkf}{\tilde{k}_{\rm f}}
\newcommand{\tku}{\tilde{k}_u}
\newcommand{\tkB}{\tilde{k}_B}
\newcommand{\chit}{\chi_{\rm t}}
\newcommand{\chiSGS}{\chi_{\rm SGS}}
\newcommand{\chiSGSz}{\chi_{\rm SGS}^{(0)}}
\newcommand{\chiSGSo}{\chi_{\rm SGS}^{(1)}}
\newcommand{\chiSGSm}{\chi^{\rm m}_{\rm SGS}}
\newcommand{\chitz}{\chi_{\rm t0}}
\newcommand{\chitm}{\overline{\chi}_{\rm t}}
\newcommand{\Co}{{\rm Co}}
\newcommand{\Cost}{\Omega_\star}
\newcommand{\Hp}{H_{\rm p}}
\newcommand{\Nu}{{\rm Nu}}
\newcommand{\Pe}{{\rm Pe}}
\newcommand{\Pra}{{\rm Pr}}
\newcommand{\PraSGS}{{\rm Pr}_{\rm SGS}}
\newcommand{\PraSGSz}{{\rm Pr}_{\rm SGS}^{(0)}}
\newcommand{\PraSGSo}{{\rm Pr}_{\rm SGS}^{(1)}}
\newcommand{\PrM}{{\rm Pr}_{\rm M}}
\newcommand{\Pm}{{\rm Pm}}
\newcommand{\Pmt}{{\rm Pm}_{\rm t}}
\newcommand{\Ra}{{\rm Ra}}
\newcommand{\Rat}{{\rm Ra}_{\rm t}}
\newcommand{\ReLS}{{\rm Re}_{\rm LS}}
\newcommand{\Rey}{{\rm Re}}
\newcommand{\Ros}{{\rm Ro}}
\newcommand{\Roc}{{\rm Ro}_{\rm c}}
\newcommand{\Rm}{{\rm Rm}}
\newcommand{\ReM}{{\rm Re}_{\rm M}}
\newcommand{\Ro}{{\rm Ro}}
\newcommand{\Sh}{{\rm Sh}}
\newcommand{\St}{{\rm St}}
\newcommand{\Lu}{{\rm Lu}}
\newcommand{\tauc}{\tau_{\rm c}}
\newcommand{\tauto}{\tau_{\rm to}}
\newcommand{\Ta}{{\rm Ta}}
\newcommand{\qij}{Q_{ij}}
\newcommand{\qxx}{Q_{xx}}
\newcommand{\qyy}{Q_{yy}}
\newcommand{\qzz}{Q_{zz}}
\newcommand{\qxy}{Q_{xy}}
\newcommand{\qxz}{Q_{xz}}
\newcommand{\qyz}{Q_{yz}}
\newcommand{\tQij}[1]{\tilde{Q}_{#1}}
\newcommand{\qrt}{Q_{r\theta}}
\newcommand{\qrp}{Q_{r\phi}}
\newcommand{\qtp}{Q_{\theta\phi}}
\newcommand{\qqij}{\mathcal{Q}_{ij}}
\newcommand{\qqrr}{\mathcal{Q}_{rr}}
\newcommand{\qqtt}{\mathcal{Q}_{\theta\theta}}
\newcommand{\qqpp}{\mathcal{Q}_{\phi\phi}}
\newcommand{\qqrt}{\mathcal{Q}_{r\theta}}
\newcommand{\qqrp}{\mathcal{Q}_{r\phi}}
\newcommand{\qqtp}{\mathcal{Q}_{\theta\phi}}
\newcommand{\qpi}{Q_{\phi i}}
\newcommand{\mij}{M_{ij}}
\newcommand{\mpi}{M_{\phi i}}
\newcommand{\mrr}{M_{rr}}
\newcommand{\mtt}{M_{\theta\theta}}
\newcommand{\mpp}{M_{\phi\phi}}
\newcommand{\mrp}{M_{r\phi}}
\newcommand{\mtp}{M_{\theta\phi}}
\newcommand{\mrt}{M_{r\theta}}
\newcommand{\mmij}{\mathcal{M}_{ij}}
\newcommand{\mmrr}{\mathcal{M}_{rr}}
\newcommand{\mmtt}{\mathcal{M}_{\theta\theta}}
\newcommand{\mmpp}{\mathcal{M}_{\phi\phi}}
\newcommand{\mmrp}{\mathcal{M}_{r\phi}}
\newcommand{\mmtp}{\mathcal{M}_{\theta\phi}}
\newcommand{\mmrt}{\mathcal{M}_{r\theta}}
\newcommand{\Rij}{R_{ij}}
\newcommand{\Rrt}{R_{r\theta}}
\newcommand{\Rrp}{R_{r\phi}}
\newcommand{\Rtp}{R_{\theta\phi}}
\newcommand{\tij}{T_{ij}}
\newcommand{\txx}{T_{xx}}
\newcommand{\tyy}{T_{yy}}
\newcommand{\tzz}{T_{zz}}
\newcommand{\txy}{T_{xy}}
\newcommand{\txz}{T_{xz}}
\newcommand{\tyz}{T_{yz}}
\newcommand{\LamV}{\Lambda_{\rm V}}
\newcommand{\LamH}{\Lambda_{\rm H}}
\newcommand{\Omx}{\Omega_x}
\newcommand{\Omz}{\Omega_z}
\newcommand{\css}{c_{\rm s}^2}
\newcommand{\emf}{\bm{\mathcal{E}}}
\newcommand{\emfi}{\mathcal{E}_i}
\newcommand{\nab}{\mbox{\boldmath $\nabla$} {}}
\newcommand{\OOO}{\hat{\mbox{\boldmath $\Omega$}} {}}
\newcommand{\meanFFFF}{\overline{\mbox{\boldmath ${\cal F}$}}{}}{}
\newcommand{\tSB}{{\rm SB}}
\newcommand{\calR}{{\cal R}}
\newcommand{\tsi}{t^{\rm sim}}
\newcommand{\usi}{U^{\rm sim}}
\newcommand{\Omsi}{\Omega^{\rm sim}}
\newcommand{\Omssi}{\Omega_\odot^{\rm sim}}
\newcommand{\rhosi}{\rho^{\rm sim}}
\newcommand{\rhosib}{\rho_{\rm bot}^{\rm sim}}
\newcommand{\rhossib}{\rho_{\odot,{\rm bot}}^{\rm sim}}
\newcommand{\Rsi}{R^{\rm sim}}
\newcommand{\Rssi}{R_\odot^{\rm sim}}
\newcommand{\Bsi}{B^{\rm sim}}
\newcommand{\gsi}{g^{\rm sim}}
\newcommand{\gssi}{g_\odot^{\rm sim}}
\newcommand{\tsu}{t_\odot}
\newcommand{\usu}{u_\odot}
\newcommand{\Omsu}{\Omega_\odot}
\newcommand{\rhosu}{\rho_\odot}
\newcommand{\rhosub}{\rho_{\odot,{\rm bot}}}
\newcommand{\Rsu}{R_\odot}
\newcommand{\Bsu}{B_\odot}
\newcommand{\gsu}{g_\odot}
\newcommand{\Lratio}{L_{\rm ratio}}
%
%
%
\newcommand{\FFFrad}{{\bm F}^{\rm rad}}
\newcommand{\FFFSGS}{{\bm F}^{\rm SGS}}
\newcommand{\FFFrads}{{\bm F}_{\rm rad}}
\newcommand{\FFFSGSs}{{\bm F}_{\rm SGS}}
\newcommand{\Fenth}{\mean{F}_{\rm enth}}
\newcommand{\Fenthr}{\mean{F}^{\rm enth}_r}
\newcommand{\Fentht}{\mean{F}^{\rm enth}_\theta}
\newcommand{\Fconv}{F_{\rm conv}}
\newcommand{\Frad}{F_{\rm rad}}
\newcommand{\Fkin}{F_{\rm kin}}
\newcommand{\Fvisc}{F_{\rm visc}}
\newcommand{\FSGS}{F_{\rm SGS}}
\newcommand{\Fbot}{F_{\rm bot}}
\newcommand{\Ftot}{F_{\rm tot}}
\newcommand{\Fenthrtp}{\brac{F^{\rm enth}_r}_{\theta \phi}}
\newcommand{\Fconvrtp}{\brac{F^{\rm conv}_r}_{\theta \phi}}
\newcommand{\Fkinrtp}{\brac{F^{\rm kin}_r}_{\theta \phi}}
\newcommand{\Fn}{\mathscr{F}_{\rm n}}
\newcommand{\mFenth}{\mean{F}_{\rm enth}}
\newcommand{\mFconv}{\mean{F}_{\rm conv}}
\newcommand{\mFrad}{\mean{F}_{\rm rad}}
\newcommand{\mFkin}{\mean{F}_{\rm kin}}
\newcommand{\mFvisc}{\mean{F}_{\rm visc}}
\newcommand{\mFSGS}{\mean{F}_{\rm SGS}}
%
%
\newcommand{\Lenthrtp}{\brac{L^{\rm enth}_r}_{\theta \phi}}
\newcommand{\Lconvrtp}{\brac{L^{\rm conv}_r}_{\theta \phi}}
\newcommand{\Lkinrtp}{\brac{L^{\rm kin}_r}_{\theta \phi}}
\newcommand{\Lconvtp}{\brac{L_{\rm conv}}_{\theta \phi}}
\newcommand{\Lenthtp}{\brac{L_{\rm enth}}_{\theta \phi}}
\newcommand{\Lkintp}{\brac{L_{\rm kin}}_{\theta \phi}}
\newcommand{\Lenthdtp}{\brac{L_{\rm enth}^{\downarrow}}_{\theta \phi}}
\newcommand{\Lconvdtp}{\brac{L_{\rm conv}^{\downarrow}}_{\theta \phi}}
\newcommand{\Lkindtp}{\brac{L_{\rm kin}^{\downarrow}}_{\theta \phi}}
\newcommand{\Lenthutp}{\brac{L_{\rm enth}^{\uparrow}}_{\theta \phi}}
\newcommand{\Lconvutp}{\brac{L_{\rm conv}^{\uparrow}}_{\theta \phi}}
\newcommand{\Lkinutp}{\brac{L_{\rm kin}^{\uparrow}}_{\theta \phi}}
\newcommand{\nabad}{\nabla_{\rm ad}}
\newcommand{\nabrad}{\nabla_{\rm rad}}
\newcommand{\namnad}{\nabla-\nabla_{\rm ad}}
\newcommand{\rhobot}{\rho_{\rm bot}}
\newcommand{\Ttop}{T_{\rm top}}
\newcommand{\Tbot}{T_{\rm bot}}
\newcommand{\tbot}{{\rm bot}}
\newcommand{\ttop}{{\rm top}}
\newcommand{\Kbot}{K_{\rm bot}}
\def\onethird{{\textstyle{1\over3}}}
\def\onehalf{{\textstyle{1\over2}}}
\def\threehalfs{{\textstyle{3\over2}}}
\def\threefourths{{\textstyle{3\over4}}}
%
\newcommand{\Figa}[1]{Fig.~\ref{#1}}
\newcommand{\Fig}[1]{figure~\ref{#1}} 
\newcommand{\Figp}[2]{figure~\ref{#1}({#2})} 
\newcommand{\Figs}[2]{figures~\ref{#1} and {#2}} 
\newcommand{\Figsp}[3]{figures~\ref{#1}({#2}) and ({#3})} 
\newcommand{\Figu}[1]{Figure~\ref{#1}}
\newcommand{\figu}[1]{figure~\ref{#1}}
\newcommand{\Seca}[1]{sect.~\ref{#1}}
\newcommand{\Sec}[1]{section~\ref{#1}} 
\newcommand{\sect}[1]{section~\ref{#1}} 
\newcommand{\Table}[1]{Table~\ref{#1}}
\newcommand{\Tablel}[1]{table~\ref{#1}}
\newcommand{\Appendix}[1]{Appendix~\ref{#1}}
\newcommand{\Appendixl}[1]{appendix~\ref{#1}}
%
%
\newcommand{\kg}{\,{\rm kg}}
\newcommand{\K}{\,{\rm K}}
\newcommand{\s}{\,{\rm s}}
\newcommand{\m}{\,{\rm m}}

\def\red{\textcolor{red}}
\def\blue{\textcolor{blue}}

%
\newcommand{\apj}{\itshape Astrophys.\ J.}
\newcommand{\apjl}{\itshape Astrophys.\ J.\ Lett.}
\newcommand{\apjs}{\itshape Astrophys.\ J.\ Suppl.}
\newcommand{\aap}{\itshape Astron.\ Astrophys.}
\newcommand{\aapr}{\itshape Astron.\ Astrophys.\ Rev.}
\newcommand{\an}{\itshape Astron.\ Nachr.}
\newcommand{\mnras}{\itshape Monthly Notices of the Roy.\ Astron.\ Soc.}
\newcommand{\pre}{\itshape Phys.\ Rev.\ E}
\newcommand{\prl}{\itshape Phys.\ Rev.\ Lett.}
\newcommand{\jfm}{\itshape J.\ Fluid Mech.}
\newcommand{\memsai}{\itshape Mem.\ d.\ Soc.\ Astron.\ It.}
\newcommand{\solphys}{\itshape Solar Phys.}
\newcommand{\zap}{\itshape Z. Astrophys.}
\newcommand{\icarus}{\itshape Icarus}
%


\title{Sensitivity to luminosity, centrifugal force, and boundary conditions in spherical shell convection}

\author{P.J.\ K\"APYL\"A$^{{\rm a,b,c,d,e}\,\ast}$,\thanks{$^\ast$Corresponding author. Email: pkaepyl@uni-goettingen.de\vspace{6pt}}
  F.A.\ GENT$^{\rm c}$,  N.\ OLSPERT$^{\rm c}$,  M.J.\ K\"APYL\"A$^{\rm d,c}$\\ and  A.\ BRANDENBURG$^{\rm e,f,g,h}$
  \\\vspace{6pt}
  $^{\rm a}$ Georg-August-Universit\"at G\"ottingen, Institut f\"ur
  Astrophysik, Friedrich-Hund-Platz 1, D-37077 G\"ottingen, Germany \\
  $^{\rm b}$Leibniz-Institut f\"ur Astrophysik, An der Sternwarte 16,
  D-14482 Potsdam, Germany \\
  $^{\rm c}$ReSoLVE Centre of Excellence, Department of Computer Science,
  P.O. Box 15400, FI-00076 Aalto, Finland \\
  $^{\rm d}$ Max-Planck-Institut f\"ur Sonnensystemforschung,
  Justus-von-Liebig-Weg 3, D-37077 G\"ottingen, Germany\\
  $^{\rm e}$NORDITA, KTH Royal Institute of Technology and Stockholm University,
  Roslagstullsbacken 23, SE-10691 Stockholm, Sweden\\
  $^{\rm f}$Department of Astronomy, AlbaNova University Center,
  Stockholm University, SE-10691 Stockholm, Sweden\\
  $^{\rm g}$JILA and Department of Astrophysical and Planetary Sciences,
  Box 440, University of Colorado, Boulder, CO 80303, USA\\
  $^{\rm h}$Laboratory for Atmospheric and Space Physics,
  3665 Discovery Drive, Boulder, CO 80303, USA\\
  \vspace{6pt}\received{\it \today,~ $ $Revision: 1.2 $ $} }

\maketitle


\begin{abstract}
We test the sensitivity of hydrodynamic and magnetohydrodynamic
turbulent convection simulations with respect to Mach number, thermal
and magnetic boundary conditions, and the centrifugal force.
We find that varying the luminosity, which also controls the Mach number,
has only a minor effect on the large-scale dynamics. A similar
conclusion can also be drawn from the comparison of two formulations
of the lower magnetic boundary condition with either vanishing
electric field or current density. The centrifugal force has an effect
on the solutions, but only if its magnitude with respect to
acceleration due to gravity is by two orders of magnitude greater than in
the Sun. Finally, we find that the parameterisation of the
photospheric physics, either by an explicit cooling term or enhanced
radiative diffusion, is more important than the thermal boundary
condition. In particular, runs with cooling tend to lead to more anisotropic
convection and stronger deviations from the Taylor-Proudman state.
In summary, the fully compressible approach taken here with the
{\sc Pencil Code} is found to be valid, while still allowing
the disparate timescales to be taken into account.
\end{abstract}


\section{Introduction}

Three-dimensional convection simulations in spherical shells are
routinely used with the aim of modelling solar and stellar differential
rotation and dynamos. Much of this work has been done with anelastic
codes such as ASH \citep[e.g.][]{BMT04}, EULAG \citep[][]{SC13},
MagIC \citep[e.g.][]{2012Icar..219..428G}, Rayleigh
\citep[e.g.][]{FH16}, and a number of unnamed codes
\citep[e.g.][]{FF14,2015ApJ...810...80S}. The main advantage of
the anelastic methods is that it is, at least in principle, possible to
use the correct solar/stellar luminosity without being severely
restricted by the acoustic time step constraint. However, the problem
of using realistic luminosity is that the thermal diffusion time
$\tau_{\rm th}$ due to the radiative conductivity becomes
prohibitively long and
simulations can typically cover only small fraction of this
\citep[e.g.][]{2017LRCA....3....1K}.

In recent years, simulations using the fully compressible
hydromagnetics equations with, e.g., the {\sc Pencil Code} \citep{BD02,B03}, have
gained popularity \citep[e.g.][]{KMB12,MK13,HRY14}. The acoustic time step issue has been dealt
with either by increasing the star's luminosity \citep[e.g.][]{KMCWB13,MMK15} or by using the
reduced sound speed technique \citep[e.g.][]{Re05,HRYIF12}, which changes the continuity equation
such that the sound speed is artificially reduced. Although the
results of fully compressible and anelastic simulations seem to
coincide \citep[][]{GYMRW14,KKOWB16}, the compromises that need to be
made in the former to model
stellar convection have not been thoroughly studied.
Here we study the effects of enhanced luminosity and caveats
associated with it. The main effect of this is the increased Mach
number which brings the dynamic and acoustic timescales closer to each
other and alleviates the time step issue \citep{KMCWB13}. While the
Mach numbers still remain clearly subsonic, this approach, however,
necessitates the use of a much higher rotation rate to reach a
comparable rotational influence as, e.g., in the Sun (see
\Appendixl{app:units} for further details). As a
consequence, the centrifugal force would be comparable to the
acceleration due to gravity and it is typically neglected
\citep[e.g.][]{KMGBC11}. Another aspect related to the increased
luminosity and rotation is that fluctuations of thermodynamic
quantities are significantly larger than in the Sun
\citep[e.g.][]{WKKB16}. This may have repercussions
for the rotation profiles via unrealistically large latitudinal
variation of temperature and turbulent heat flux.

Common to all of the numerical simulations of stellar convection is
the use of a wide selection of thermal and magnetic boundary conditions (BCs).
In stars the convection zones are delimited by radiative and
coronal layers without sharp boundaries. Although it is becoming
possible to include such layers self-consistently in global spherical
models \citep[][]{BMT11,WKMB13,GSdGDPKM15}, such models
necessarily have lower spatial resolution or require exceptional
computational resources. Thus the majority of present simulations
still consider only the convection zone where BCs come into play. The
BCs are typically compromises between physical accuracy and numerical
convenience. Often the implicit assumption is that the
BCs play only a minor role for the solutions. However,
this is another aspect that has not been well studied.

Here we set out to study a subset of the issues raised above. More
specifically, we use the {\sc Pencil Code} to study the sensitivity of
hydrodynamic (HD) and magnetohydrodynamic (MHD) simulations to changes in the
luminosity, to adopting subsets of typical BCs used in the literature, and to varying the
centrifugal force.

\section{Model}

\subsection{Basic equations and their treatment}

Our simulation setup is similar to that used in
\cite{2018arXiv180305898K} with a few variations that will be
explained in detail. We solve a set of fully compressible
hydromagnetics equations
\begin{align}
\frac{\pd {\bm A}}{\pd t}\, = \,&\,{\bm U} \times {\bm B} - \eta \mu_0 {\bm J}\,, \\
\frac{\DR \ln\rho}{\DR t}\, =\,& \,- \,\bm\nabla\bm\cdot{\bm U}\,, \\
\frac{\DR {\bm U}}{\DR t}\, =\, &\,\msFFF^{\rm grav} + \msFFF^{\rm Cor} + \msFFF^{\rm cent} - \frac{1}{\rho}(\bm\nabla p + {\bm J} \times {\bm B} + \bm\nabla \bm\cdot 2\nu\rho\bm{\mathsf{S}})\,,\\
T\frac{\DR s}{\DR t} \,= \,&\, \frac{1}{\rho} \left[ \eta\mu_0 {\bm J}^2 - \bm\nabla\bm\cdot( {\bm F}^{\rm rad} + {\bm F}^{\rm SGS} ) - \Gamma_{\rm cool} \right] + 2\nu \bm{\mathsf{S}}^2\,, \label{equ:ss}
\end{align}
where ${\bm A}$ is the magnetic vector potential, ${\bm U}$ is the
velocity, ${\bm B} = \bm\nabla\times{\bm A}$ is the magnetic field,
$\eta$ is the magnetic diffusivity, $\mu_0$ is the permeability of
vacuum, ${\bm J}=\bm\nabla\times{\bm B}/\mu_0$ is the current density,
$\DR/\DR t = \pd/\pd t + {\bm U}\bm\cdot\bm\nabla$ is the advective time
derivative, $\rho$ is the density, $\nu$ is
the kinematic viscosity, $p$ is the pressure, and $s$ is the specific
entropy with $\DR s=\cV \DR \ln p-\cP \DR \ln\rho$, where $\cV$ and $\cP$ are
the specific heats at constant volume and pressure, respectively. The
gas is assumed to obey the ideal gas law, $p=\mathcal{R}\rho T$, where
$\mathcal{R}=\cP-\cV$ is the gas constant. The rate of strain tensor
is given by
\begin{eqnarray}
\mathsf{S}_{ij} \,=\, \onehalf (U_{i;j} + U_{j;i}) - \onethird \delta_{ij} \bm\nabla\bm\cdot {\bm U}\,,
\end{eqnarray}
where the semicolons refer to covariant derivatives
\citep{MTBM09}. The acceleration due to gravity, and the Coriolis and
centrifugal forces are given by
\begin{align}
\msFFF^{\rm grav}\, =\,&\, -\, \big({GM_\odot}\big/{r^2}\bigr) \hat{\bm r} \,\equiv\, {\bm g}\,,  \\
\msFFF^{\rm Cor} \, =\,&\, -\, 2\bm\Omega_0\times{\bm U}\,, \\ \label{equ:centri}
\msFFF^{\rm cent} \,= \,&\,-\, c_{\rm cent}\bm\Omega_0 \times (\bm\Omega_0 \times {\bm r})\,,
\end{align}
where $G=6.67\cdot10^{-11}$~N~m$^2$~kg$^{-2}$ is the universal
gravitational constant, $M_\odot=2.0\cdot10^{30}$~kg is the solar
mass, $\bm\Omega_0=(\cos\theta,-\sin\theta,0)\Omega_0$ is the angular
velocity vector, where $\Omega_0$ is the rotation rate of the frame of
reference, $\rrr$ is the radial coordinate, and $\hat{\bm
  r}=\rrr/|\rrr|$ the corresponding radial unit vector. The parameter
$c_{\rm cent}$ is used to control the magnitude of the centrifugal
force.

Radiation is taken into account via a diffusive radiative flux
\begin{eqnarray}
{\bm F}^{\rm rad} \,=\, -\,K\bm\nabla T\,,
\label{equ:Frad}
\end{eqnarray}
where $K=\cP \rho \chi$ is the heat conductivity. Here $K$ has either
a fixed profile as a function of radius $K=K(r)$ or it is a function of
density and temperature $K=K(\rho,T)$. In the former case we use the
profile defined in \cite{KMCWB13}. In the latter case $K$ adapts
dynamically with the thermodynamic state and is computed from
\begin{eqnarray}
K \,= \,\frac{16 \sigma_{\rm SB} T^3}{3 \kappa \rho}\,,
\label{equ:Krad1}
\end{eqnarray}
where $\sigma_{\rm SB}$ and $\kappa$ are the Stefan-Boltzmann constant
and opacity, respectively. For the latter a power law as a function of
$\rho$ and $T$ is assumed
\begin{eqnarray}
\kappa \,=\, \kappa_0 (\rho/\rho_0)^a (T/T_0)^b\,,
\label{equ:kappa}
\end{eqnarray}
where $\rho_0$ and $T_0$ are reference values of density and
temperature. Here these quantities are the values of $\rho$ and $T$
from the initially non-convecting state at the bottom of the
domain. Equations~(\ref{equ:Krad1}) and (\ref{equ:kappa}) yield
\citep{BB14}
\begin{eqnarray}
K(\rho,T)\, =\, K_0 (\rho/\rho_0)^{-(a+1)} (T/T_0)^{3-b}\,.
\label{equ:Krad2}
\end{eqnarray}
Here we use $a=1$ and $b=-7/2$, corresponding to the Kramers opacity
law for free-free and bound-free transitions \citep{WHTR04}. This
formulation has previously been used in local
\citep{2000gac..conf...85B,2017ApJ...845L..23K} and semi-global
\citep{2018arXiv180305898K} simulations of convection.
We refer to the heat conductivity introduced in \Equ{equ:Krad2} as
$K^{\rm Kramers}$.
Here we also consider a few cases where a fixed profile of $K$ is
used near the surface -- in addition to the Kramers conductivity. In such
cases the value of $K$ near the surface is artificially enhanced, and
denoted $K^{\rm surf}$, to
facilitate the outwards transport of thermal energy. This can be
considered a crude parameterisation of the effective radiative
transport in the photosphere.

The thermal diffusivity from the radiative conductivity, $\chi=K/\cP
\rho$, can vary by
several orders of magnitude as a function of radius which can lead to
numerical instability. Thus, an additional subgrid scale (SGS)
diffusion is applied in the entropy equation:
\begin{eqnarray}
{\bm F}^{\rm SGS}\, =\, -\,\chiSGS \rho T \bm\nabla s'\,,
\label{equ:FSGS}
\end{eqnarray}
where $\chiSGS$ is the (constant) SGS diffusion coefficient. The SGS
diffusion acts on fluctuations of entropy
$s'(r,\theta,\phi)=s-\brac{s}_{\theta\phi}$, where $\brac{s}_{\theta\phi}$ is the
horizontally averaged or spherically symmetric part of the specific
entropy.

The penultimate term on the right-hand side of (\ref{equ:ss}) models radiative
cooling near the surface of the star:
\begin{eqnarray}
  \Gamma_{\rm cool} \,= \,-\,\Gamma_0 f(r) (T_{\rm cool}-\brac{T}_{\theta \phi})\,,
\label{equ:Gamma}
\end{eqnarray}
where $\Gamma_0$ is a cooling luminosity, $\brac{T}_{\theta \phi}$ is
the spherically symmetric part of the temperature, and $T_{\rm
  cool}=T_{\rm cool}(r)$ is a radially varying reference temperature
coinciding with the initial stratification. We use the {\sc Pencil
  Code}\footnote{\url{https://github.com/pencil-code/}},
which uses sixth order finite differences in its standard configuration
and a third-order accurate time-stepping scheme.
Curvilinear coordinates are implemented by replacing derivatives by
covariant ones; see appendix~B of \cite{MTBM09}.

\subsection{System parameters and diagnostics quantities}

The simulations were done in spherical wedges with $r_0<r<R_\odot$,
where $r_0=0.7R_\odot$ and
$R_\odot=7\cdot10^8$~m is the solar radius,
$15^\circ<\theta<165^\circ$ in colatitude, and $0<\phi<90^\circ$ in
longitude. The simulations are fully defined by specifying the energy
flux imposed at the bottom boundary, $F_{\rm bot}=-(K \pd T/\pd
r)|_{r=r_0}$, the values of $K_0$, $a$, $b$, $\rho_0$, $T_0$,
$\Omega_0$, $\nu$, $\eta$, $\chiSGS$, and the fixed profile of $K$ in
cases where a fixed profile of $K$ is used. Finally, the profile of
$f(r)$ is piecewise constant with $f(r)=0$ in $r_0 < r
<0.99R_\odot$, and connecting smoothly to $f(r)=1$ above
$r=0.99R_\odot$.

Due to the fully compressible formulation used in the current
simulations, we use a much higher luminosity than in the target star to avoid
the time step being limited by sound waves. This also necessitates the
use of a much higher rotation rate to reach an equivalent rotational
state as in the target star. This leads to a situation where the results need
to be scaled accordingly to represent them in physical units, see
\Appendixl{app:units}.

The parameters describing the simulations include the non-dimensional
luminosity
\begin{eqnarray}
\mathcal{L}\,=\,\frac{L_0}{\rho_0 (GM_\odot)^{3/2}R_\odot^{1/2}}\,,
\end{eqnarray}
the non-dimensional pressure scale height at the surface controlling
the initial stratification
\begin{eqnarray}
\xi_0\,=\,\frac{\mathcal{R}T_1}{GM_\odot/R_\odot}\,,
\end{eqnarray}
where $T_1$ is the temperature at the surface ($r=R_\odot$).

The Prandtl numbers describing the ratios between viscosity, SGS
diffusion, and magnetic diffusivity are given by
\begin{eqnarray}
\PraSGS\, =\,{\nu}\big/{\chiSGS}\,, \hskip 20mm \Pm={\nu}\big/{\eta}\,.
\end{eqnarray}
$\PraSGS=\Pm=1$ in all of our runs. The thermal Prandtl number
associated with the radiative diffusivity is
\begin{eqnarray}
\Pr\, = \,{\nu}\big/{\chi}\,.
\end{eqnarray}
In distinction to $\PraSGS$ and $\Pm$, $\Pr$ in general varies as a
function of radius and time, especially in cases where the Kramers opacity is
used.

The efficiency of convection is traditionally given in terms of the
Rayleigh number computed from the non-convecting, hydrostatic state:
\begin{eqnarray}
\Ra\, = \,\frac{GM_\odot (\Delta r)^4}{\nu \chiSGS R_\odot^2}\left(- \frac{1}{c_{\rm P}} \frac{{\rm d}s_{\rm hs}}{{\rm d}r}\right)_{r_{\rm m}},
\end{eqnarray}
where $\Delta r=0.3R_\odot$ is the depth of the layer, $s_{\rm hs}$ is
the specific entropy, evaluated at the middle of the domain at $r_{\rm
  m}=0.85R_\odot$. With the Kramers-based heat conduction prescription
the convectively unstable layer in the hydrostatic state is confined
to a thin surface layer see, e.g., figure~7 of \cite{Br16}. Thus $\Ra<0$
at $r=r_{\rm m}$, rendering this definition irrelevant for the current
simulations. It is, however, possible to define a `turbulent' Rayleigh
number ($\Rat$) where the actual entropy gradient ${\rm d}s/{\rm d}r$
from the thermally saturated state
is used instead of the hydrostatic one
\citep[e.g.][]{KMCWB13,2018ApJ...859..117N}.

Furthermore, we also quote the Nusselt number
\citep[e.g.][]{HTM84,Br16}:
\begin{eqnarray}
\Nu\, =\, {\nabrad}\big/{\nabad}\,,
\end{eqnarray}
near the surface at $r=0.98R_\odot$ where
\begin{equation}
\nabrad\,=\,\frac{\mathcal{R}}{Kg}\Ftot\,,  \hskip 12mm \mbox{and} \hskip 12mm \nabad\,=\,1-\frac{1}{\gamma}\,,
\end{equation}
are the radiative and adiabatic temperature gradients, and where
$g=|{\bm g}|$, and $\Ftot=L_0/(4\pi r^2)$.

The strength of rotation is given in terms of the Taylor number
\begin{eqnarray}
\Ta\, =\, \bigl(2\Omega_0 \Delta r^2/\nu^2\bigr)^2\,.
\end{eqnarray}

The remaining quantities are used as diagnostics and they are based on
the outcomes of the simulations. The fluid and magnetic Reynolds
numbers quantify the influence of the applied diffusion coefficients,
and are given by
\begin{eqnarray}
\Rey\,=\,\frac{\Urms}{\nu k_1}\hskip 12mm \mbox{and} \hskip 12mm
\ReM\,=\,\frac{\Urms}{\eta k_1}\,,
\end{eqnarray}
respectively, where $\Urms$ is the rms value of the total velocity,
and $k_1=2\pi/\Delta r\approx21/R_\odot$ is the
wavenumber corresponding to the depth of the domain.

The Coriolis number quantifies the rotational influence on the flow
\begin{eqnarray}
\Co\,=\,\frac{2\,\Omega_0}{\Urms k_1}\,.
\end{eqnarray}
Mean quantities refer to azimuthal (denoted by an overbar) or
horizontal averages (denoted by angle brackets with subscript
$\theta\phi$). In addition, time averaging is also performed unless
explicitly stated otherwise.

\subsection{Initial and boundary conditions}

The majority of the simulations presented here are based on Run~RHD2 of
\cite{2018arXiv180305898K}. The initial stratification is isentropic,
described by a polytropic index of $n=1.5$. The initial density
contrast of roughly 80 which results in from the choice of
$\xi_0=0.01$. In the initial state the radiative flux is very small in
the upper part of the domain and the system is thus not in
thermodynamic equilibrium. Convection is driven by the efficient
surface cooling \citep[see e.g.][]{KMCWB13}. The value of $K_0$ in the
models with Kramer-based heat conduction is chosen such that a stably
stratified overshoot layer of extent $d_{\rm os}\approx0.05R_\odot$
develops at the base of the domain. In cases with a fixed heat
conductivity profile, the value of $K$ at $r=r_0$ is set such
that the flux through the boundary is $L_0/4\pi r_0^2$.

The following BCs are common to all runs: the radial
and latitudinal boundaries are assumed impenetrable and stress-free
for the flow
\begin{align}
 U_r\,=\,&\,0\,, \hskip 15mm \frac{\pd U_\theta}{\pd r}\,=\,\frac{U_\theta}{r}\,,\quad &&\frac{\pd U_\phi}{\pd r}\,=\,\frac{U_\phi}{r}
& (r=r_0, R_\odot)&\,,\\
\frac{\pd U_r}{\pd \theta}\,=\,&\,U_\theta\,=\,0\,,\quad &&\frac{\pd U_\phi}{\pd \theta}\,=\,U_\phi \cot \theta
& (\theta=\theta_0,\pi-\theta_0)&\,.\hskip 10mm
\end{align}
On the bottom boundary, a fixed heat flux is prescribed:
\begin{eqnarray}
\Fbot\, =\, -\, \Kbot(\theta,\phi) \frac{\pd T}{\pd z}\hskip 10mm  (r=r_0)\,,
\end{eqnarray}
where we have emphasised that $\Kbot$ is in general nonuniform.
On the latitudinal boundaries, the gradients of thermodynamic
quantities are set to zero
\begin{eqnarray}
\frac{\pd s}{\pd \theta}\, =\, \frac{\pd \rho}{\pd \theta}\, =\,0
\hskip 10mm (\theta=\theta_0,\pi-\theta_0)\,.
\label{equ:dsdrho}
\end{eqnarray}
Although there is no BC on $\rho$, we impose
\equ{equ:dsdrho} as a symmetry condition to populate the ghost zones
in the numerical calculations.
Finally, the magnetic field in the MHD runs is radial at the outer
boundary and tangential on the latitudinal boundaries, which
translate to
\begin{align}
A_r \,=\,& \,0\,, \hskip 12mm \frac{\pd A_\theta}{\pd r}\, =\, -\,\frac{A_\theta}{r}\,,\hskip 12mm \frac{\pd A_\phi}{\pd r}\, =\, -\,\frac{A_\phi}{r} & (r=R_\odot)&\,,\\
A_r \,=\,&\, \frac{\pd A_\theta}{\pd \theta}\, = \,A_\phi\, =\, 0 & (\theta=\theta_0,\pi-\theta_0)&\,,\hskip 10mm
\end{align}
in terms of the magnetic vector potential.

The following conditions are varied in the simulations. The upper
thermal boundary is chosen from three possibilities:
\begin{align}
T \,=\,&\, {\rm const.} \hspace{1.1cm} \mbox{(cT)}\,, \label{equ:cT} \\
F_r^{\rm rad}\, =\,& \,\sigma T^4 \hspace{1.42cm} \mbox{(bb)}\,, \label{equ:bb} \\
\frac{\pd s}{\pd r} =\,& 0 \hspace{1.95cm} \mbox{(ds)}\,, \label{equ:ds}
\end{align}
which correspond to constant temperature (cT), black body (bb), and
vanishing radial derivative of entropy (ds) and where $\sigma$ is a
modified Stefan--Boltzmann constant. For the magnetic field at
the lower boundary $(r=r_0)$ we either assume vanishing tangential
electric field (vE) or additionally vanishing tangential
currents (vJ):
\begin{align}
\frac{\pd A_r}{\pd r} \,= \,&\,A_\theta\,=\,A_\phi \,=\,0 & (\mbox{vE and vJ})\,,& \label{equ:vE} \\
\frac{\pd^2 A_\theta}{\pd r^2} \,= \,&-\,\frac{2}{r_0}\frac{\pd A_\theta}{\pd r}\,,  \hskip 12mm \frac{\pd^2 A_\phi}{\pd r^2}\,=\, -\,\frac{2}{r_0}\frac{\pd A_\phi}{\pd r} & (\mbox{vJ})\,.&
\end{align}
Note that for the vJ conditions both equations must be
fulfilled. The azimuthal direction is periodic for all quantities.

The initial conditions for the velocity and magnetic fields are random
Gaussian noise fluctuations with amplitudes on the order of
$0.1$~m~s$^{-1}$ and $0.1$~Gauss, respectively.

\begin{table}
  \tbl{Summary of the input parameters runs. All runs have $\PraSGS=1$ and
    grid resolution $144\times288\times144$.}
{\begin{tabular}{@{}lcccccccccccc}\toprule
    Run & $\mathcal{L} [10^{-6}]$ & $\Lratio [10^{5}]$ & $\tilde{\Omega}$ & $c_{\rm cent} [10^{-2}]$ & $\Ta [10^7]$ & $\xi_0$ & $\PrM$ & Surf. & $\tilde{\Gamma}_0$ & $\tilde{\sigma}[10^3]$ & BCt & BCm \\ \colrule
    A1     &  10  &  2.1 &  3 & 0 & 2.3 & 0.01 & -- & cool &  1/3 & -- & cT &  -- \\
    A2     &   5  &  1.1 &  3 & 0 & 2.3 & 0.01 & -- & cool &  1/6 & -- & cT &  -- \\
    A3     &   2  &  0.4 &  3 & 0 & 2.3 & 0.01 & -- & cool & 1/15 & -- & cT &  -- \\
    A4     &   1  &  0.2 &  3 & 0 & 2.3 & 0.01 & -- & cool & 1/30 & -- & cT &  -- \\
    \hline
    A2c1   &   5  &  1.1 &  3 &  0.05 & 2.3 & 0.01 & -- & cool & 1/6 & -- & cT &  -- \\
    A2c2   &   5  &  1.1 &  3 &  0.5  & 2.3 & 0.01 & -- & cool & 1/6 & -- & cT &  -- \\
    A2c3   &   5  &  1.1 &  3 &   5   & 2.3 & 0.01 & -- & cool & 1/6 & -- & cT &  -- \\
    \hline
    A4bb   &   1  &  2.1 &  3 & 0 & 2.3 & 0.01 & -- & diff ($K$) &  --  & 18 & bb & -- \\
    A4ds   &   1  &  2.1 &  3 & 0 & 2.3 & 0.01 & -- & cool       & 1/30 & -- & ds & -- \\
    A4ds2  &   1  &  2.1 &  3 & 0 & 2.3 & 0.01 & -- & diff ($K$) &  --  & 18 & ds & -- \\
    \hline
    M1     &  38  &  13  &  5 & 0 & 12 & 0.02 & 1.0 & diff ($\chit$) & -- & 1.4 & bb & vE \\ 
    M2     &  38  &  13  &  5 & 0 & 12 & 0.02 & 1.0 & diff ($\chit$) & -- & 1.4 & bb & vJ \\ 
    \botrule
  \end{tabular}}
\tabnote{The photospheric layers are parameterised through cooling (cool),
  diffusion (diff) due to radiative heat conductivity ($K$) or subgrid
  scale turbulent entropy diffusion ($\chit$). For the latter, see
  \cite{KMCWB13}. Furthermore, $\tilde{\Gamma}_0 = \Gamma_0 (GM)^{1/2} /
  \rho_0 \cP R_\odot^{3/2}$ and $\tilde{\sigma}=\sigma
  R_\odot^2 T_0^4/L_0$ where $\rho_0$ and $T_0$ are the density and
  temperature at $r_0=0.7R_\odot$ in the initial non-convecting
  state.}
\label{tab:runs}
\end{table}

\begin{table}
  \tbl{Summary of the diagnostic quantities.}
{\begin{tabular}{@{}lccccccccc}\toprule
    Run & $\Rat [10^5]$ & $\Nu_0 [10^3]$ & $\Nu [10^3]$ & $\Rey$ & $\ReM$ & $\Co$ & $\Delta \rho_0$ & $\Delta \rho$ & $\Delta t\ \mbox{[yr]}$ \\ \colrule
    A1     &  7.1  &   4.1  &  4.0  &  31  &  --  &  4.0  &  77  &  62  & 28 \\
    A2     &  7.4  &   4.1  &  3.9  &  31  &  --  &  3.9  &  77  &  67  &  8 \\
    A3     &  7.8  &   4.1  &  3.9  &  31  &  --  &  3.9  &  77  &  71  & 13 \\
    A4     &  7.9  &   4.1  &  4.0  &  32  &  --  &  3.9  &  77  &  73  & 14 \\
    \hline
    A2c1   &  7.3  &   4.1  &  3.9  &  31  &  --  &  3.9  &  77  &  67  & 14 \\
    A2c2   &  7.4  &   4.1  &  3.9  &  31  &  --  &  3.9  &  77  &  66  & 15 \\
    A2c3   &  6.8  &   4.1  &  3.8  &  30  &  --  &  4.1  &  77  &  62  & 15 \\
    \hline
    A4bb   &  9.8  &  0.045 & 0.045 &  33  &  --  &  3.7  &  77  & 111  & 12 \\
    A4ds   &  8.1  &   4.1  &  4.0  &  32  &  --  &  3.9  &  77  &  73  & 13 \\
    (A4ds2  &  10.1 &  0.045 & 0.045 &  35  &  --  &  3.5  &  77  & 108  & 21) \\
    \hline
    M1     &  2.8  &  0.32  &  0.32 &  29  &  29  &  9.5  &  30  &  19  & 45 \\ 
    M2     &  2.8  &  0.32  &  0.32 &  29  &  29  &  9.5  &  30  &  19  & 45 \\ 
    \botrule
  \end{tabular}}
\tabnote{$\Nu_0$ and $\Nu$ are the Nusselt numbers from the initial
  and saturated stages, respectively. $\Delta t$ gives the length of
  the saturated stage of the simulations in years. Run~A4ds2 is
  included for completeness although it does not reach a relaxed state
  in the time ran here, see \sect{sec:thermalBCs}.}
\label{tab:diagnostics}
\end{table}

\section{Results}

We perform four sets of simulations where different aspects of the
model are varied.
These include changing the luminosity, centrifugal force,
and thermal or magnetic BCs. For the first three HD sets we
use run~RHD2 of \cite{2018arXiv180305898K} as progenitor run, which is the
same as our Run~A1. Runs~A[2-4] were then branched off from this model by
changing the luminosity, diffusion coefficients, and cooling
luminosity in the initial state. Runs~A2c[1-3] (A4[bb,ds,ds2]) were
run from the same initial conditions as run~A2 (A4). In the
last MHD set, the `millennium' run of
M.\ \cite{KKOBWKP16} and the run
presented in \cite{GKW17} are denoted as Runs~M1 and M2,
respectively. The input parameters of the runs are listed in
\Tablel{tab:runs}.

\begin{figure}
\begin{center}
    \includegraphics[width=.5\textwidth]{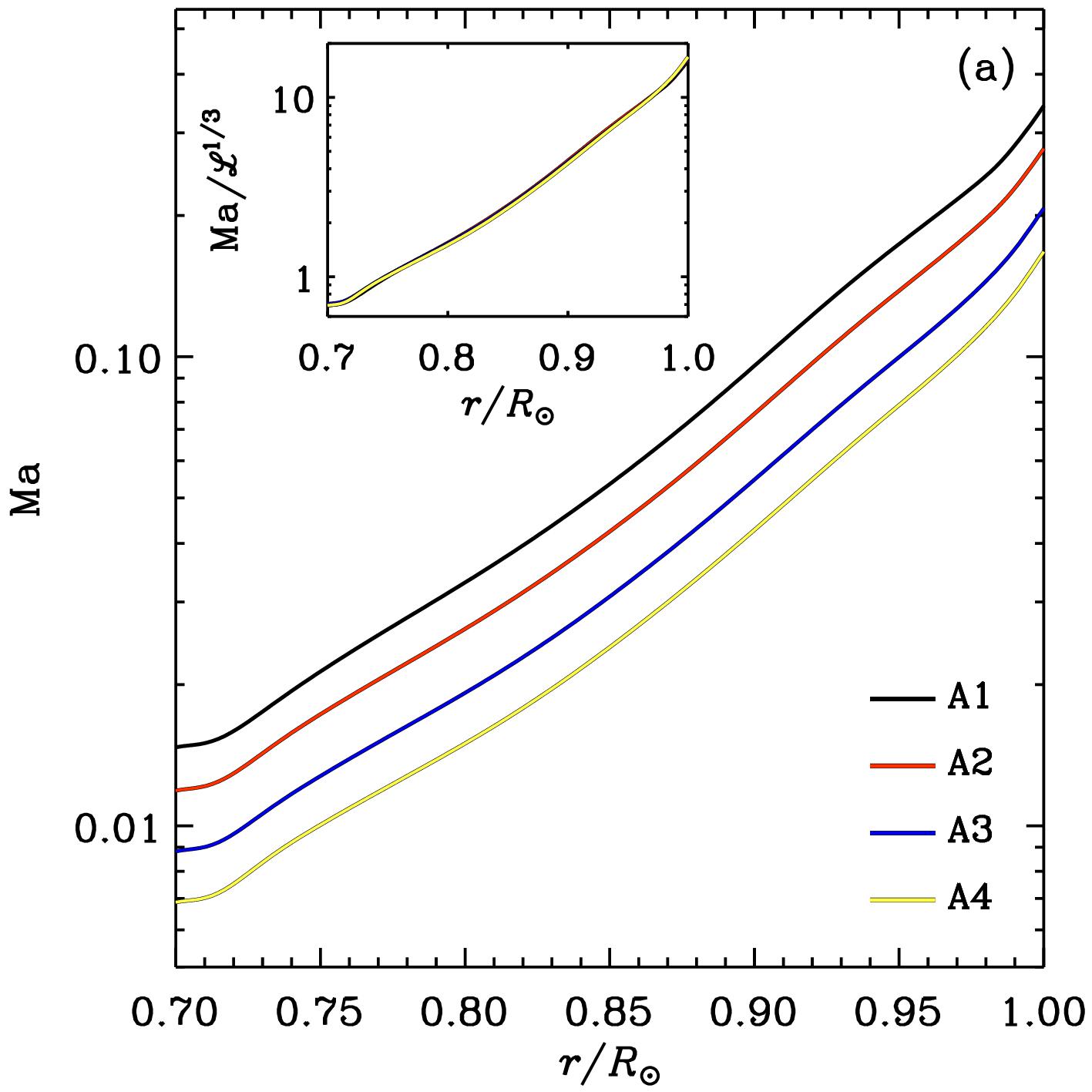}\includegraphics[width=.5\textwidth]{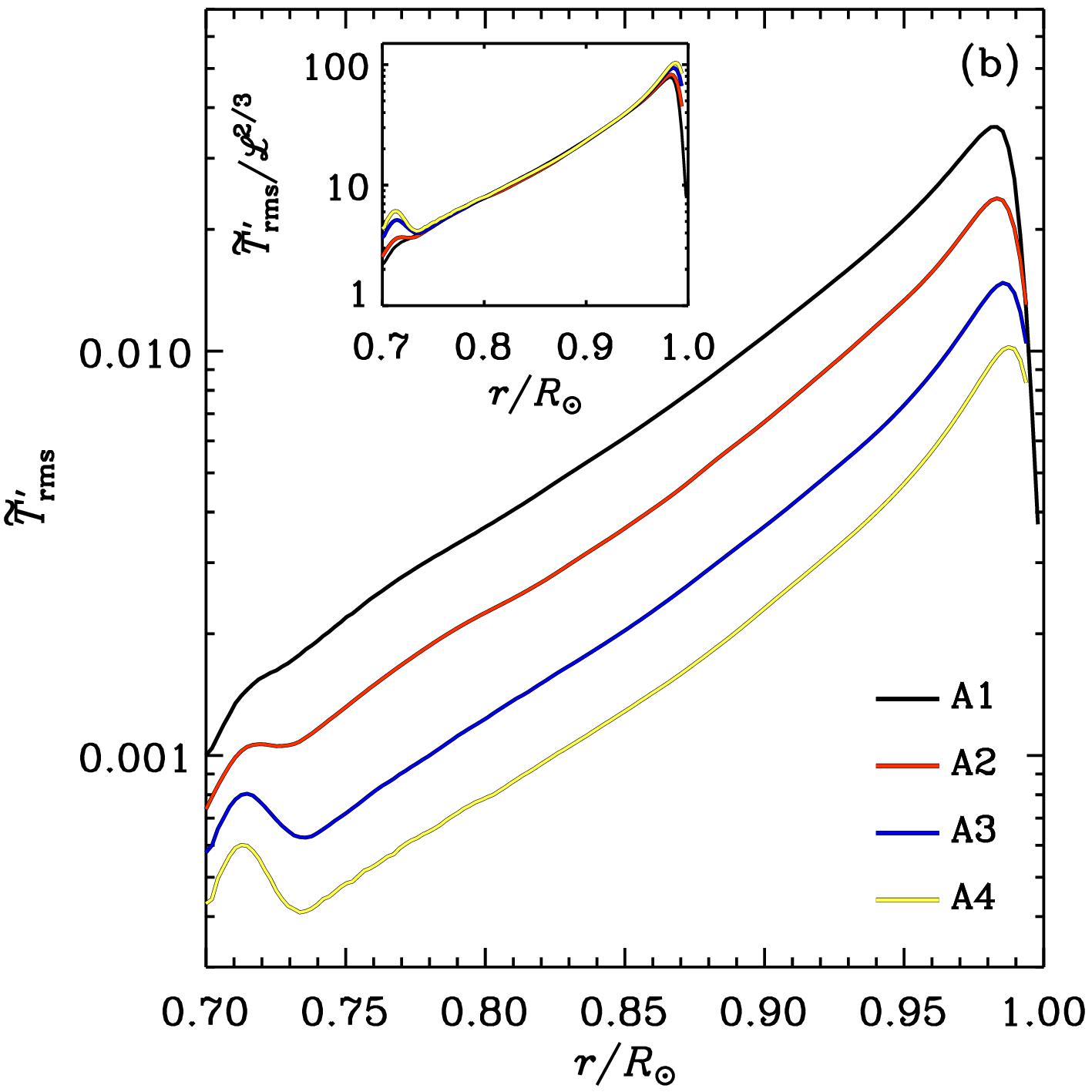}
    \caption{a) Horizontally averaged Mach number as a function of
      radius from Runs~A1--A4. The inset shows the Mach numbers
      normalised by $\mathcal{L}^{1/3}$. b) Horizontally averaged
      normalised rms temperature fluctuation $\tilde{T}'_{\rm rms} =
      T'_{\rm rms}/\brac{T}_{\theta\phi}$ as a function of $r$ from
      the same runs. The inset shows $\tilde{T}'_{\rm rms}$ normalised
      by $\mathcal{L}^{2/3}$ (colour online).}
\label{fig:pMach}
\end{center}
\end{figure}

\subsection{Varying luminosity}

One of the disadvantages of solving the fully compressible equations
is that if a realistic luminosity for the star is used, the flow
velocities are much smaller than the sound speed, with
the latter imposing a prohibitively short time step. In the case of
the {\sc Pencil Code} this has been circumvented by enhancing the
luminosity by a factor that is typically on the order of
$10^5\ldots10^6$ \citep[e.g.][]{KKB14,2018arXiv180305898K}. The
luminosity enhancement procedure and the way how to relate the model results
to physical units is discussed in detail in \Appendixl{app:units}. The
ratio of the dimensionless luminosities in the simulations in
comparison to the Sun quantifies this procedure:
\begin{eqnarray}
\Lratio\, = \, \mathcal{L}/\mathcal{L}_\odot\, .
\end{eqnarray}
Values of $\Lratio$ quoted above are sufficiently high to decrease the
thermal diffusion time such that it is possible to fully thermally
relax the simulations \citep{KMCWB13}. The downside is that the
velocity as well as the fluctuations of thermodynamic quantities are
unrealistically high \citep{WKKB16}.
It has been speculated that such effects contribute to features such as
convectively stable regions at certain mid-latitudes
\citep[e.g.][]{KMGBC11,2018arXiv180305898K}.
Here we vary the luminosity by one order of magnitude in
Runs~A1--A4; see \Tablel{tab:runs}. To isolate the effects of the
luminosity we keep the
Reynolds and Coriolis numbers fixed by varying the viscosity $\nu$ and
rotation rate of the frame $\Omega_0$ with
$\mathcal{L}^{1/3}$, see \Appendixl{app:units} and
\Tablel{tab:diagnostics}. Similarly the cooling luminosity is varied
with a 1/3 power of $\mathcal{L}$.

We examine first the scaling of convective velocity and temperature
fluctuations as function of the luminosity.
The horizontally and temporally averaged Mach number,
$\Ma=\Urms(r)/\cs$, is shown in \Fig{fig:pMach}(a). $\Ma$ decreases
monotonically as $\mathcal{L}$ is decreased. The inset shows
that the convective velocity scales with the $1/3$ power of the
luminosity. Furthermore, the horizontally and temporally averaged rms
value of the temperature fluctuation $T'_{\rm
  rms}(r)=\sqrt{\brac{T'^2}_{\theta\phi}}$, where $T'=T-\mean{T}$,
also shows a
decrease with $\mathcal{L}$, and a scales with $2/3$ power of
$\mathcal{L}$. Both results agree with the expected behaviour from
mixing length arguments \citep[][]{BCNS05}.

\begin{figure}
\begin{center}
\begin{minipage}{150mm}
\subfigure[]{
\resizebox*{7cm}{!}{\includegraphics{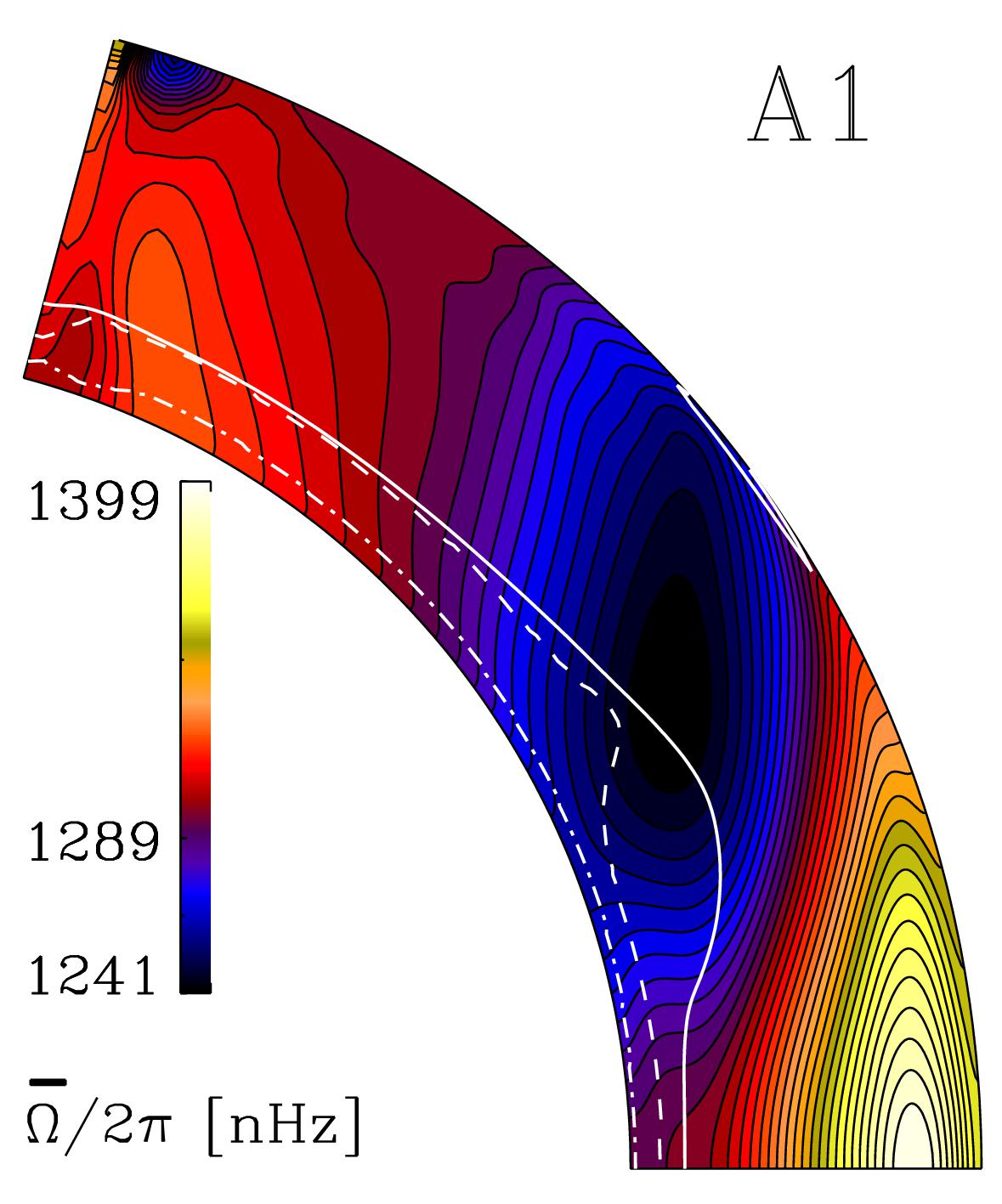}}}%
\subfigure[]{
\resizebox*{8cm}{!}{\includegraphics{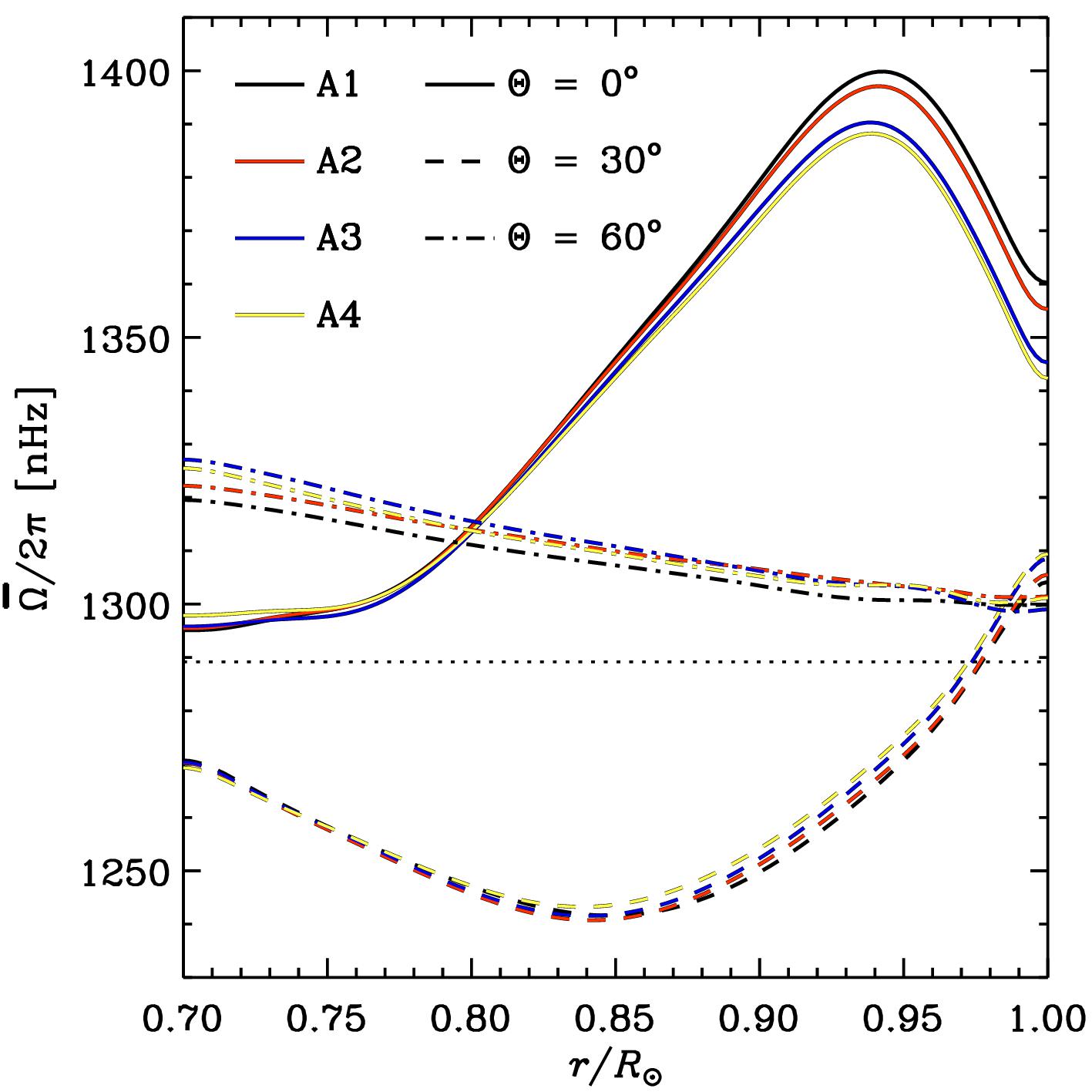}}}%
\caption{(a) Temporally averaged mean angular velocity
  $\mean{\Omega}=\mean{U}_\phi/r \sin\theta + \Omega_0$ from
  Run~A1. The white solid, dashed, and dot-dashed lines denote the
  bottoms of the BZ, DZ, and OZ, respectively. (b) $\mean{\Omega}$
  from latitudes $0^\circ$ (solid lines),
  $30^\circ$ (dashed), and $60^\circ$ (dash-dotted) for Runs~A1
  (black), A2 (red), A3 (blue), and A4 (yellow) (colour online).}
\label{fig:pOm_lumi}
\end{minipage}
\end{center}
\end{figure}

The mean angular velocity profile $\mean{\Omega}=\mean{U}_\phi/r \sin
\theta + \Omega_0$ from Run~A1 is shown in \figu{fig:pOm_lumi}(a). The
rotation profile is solar-like with a fast equator, but a prominent
mid-latitude minimum is also present. This is a common feature in
many current simulations
\citep[e.g.][]{KMB11,MMK15,ABMT15,2018ApJ...859...61B} and it is the
most likely cause of the equatorward
migrating large-scale magnetism observed in several MHD models of
solar-like stars \citep{WKKB14}. \Figu{fig:pOm_lumi}(b) shows the
radial profiles of
$\mean{\Omega}$ from three latitudes from Runs~A1--A4. We find that
the rotation profiles in these runs are very similar, with the only
consistent trend being the weakly decreasing equatorial rotation rate
as a function of $\mathcal{L}$. Thus the Mach number has only a weak
effect on the large-scale flows in the parameter range studied here.

We use the nomenclature introduced in
\cite{2017ApJ...845L..23K,2018arXiv180305898K} to classify the
different radial layers in the system
\citep[see also][]{2015ApJ...799..142T}. This classification depends on
the signs of the radial enthalpy flux $\Fenthr = \cP \mean{(\rho U_r)'
  T'}$ and the radial
gradient of specific entropy, $\nabla_r \mean{s} = \pd \mean{s}/\pd r$. The
buoyancy zone (BZ) is characterized by $\nabla_r \mean{s}<0$ and
$\Fenthr>0$, whereas in the Deardorff zone (DZ), $\nabla_r \mean{s}>0$
and $\Fenthr>0$. Here, as emphasised by \cite{Br16} in the astrophysical
context, the outward enthalpy flux can only be carried by Deardorff's
non-gradient contribution; see \cite{De66}.
Finally, in the overshoot zone (OZ), $\Fenthr<0$ and
$\nabla_r\mean{s}>0$, and its bottom is located where $|\Fenthr|$
falls below a threshold value, here chosen to be $0.025L_0$.
\Figu{fig:pOm_lumi}(a) also shows the lower boundaries of the
buoyancy, Deardorff, and overshoot zones in Run~A1. We do not find a
significant variation of the depths of the zones in the studied range
of $\mathcal{L}$. Furthermore, a radiation zone where $|\Fenthr|\approx 0$ and
$\mFrad\approx\Ftot$, does not have room to
develop in these runs and the overshoot layer tends to extend all the
way to the lower boundary of the domain. Thus, it is not possible to
draw conclusions
about the scaling of the overshoot depth as a function of luminosity
\citep[e.g.][]{1998A&A...340..178S,2009MNRAS.398.1011T,2017ApJ...843...52H}.

The contours of angular velocity are clearly inclined with respect to
the rotation vector in Runs~A1--A4, which indicates deviation from the
Taylor-Proudman balance. To study this, we consider the vorticity
equation in the meridional plane:
\begin{eqnarray}
\frac{\pd \mean{\omega}_\phi}{\pd t} \,= \,r \sin \theta \frac{\pd \mean{\Omega}^2}{\pd z} + (\bm\nabla \mean{T} \times \bm\nabla \mean{s})_\phi + \cdots\,, \label{equ:vortphi}
\end{eqnarray}
where $\mean{\bm\omega} = \bm\nabla \times \mUUU$, and where
$\pd/\pd z = \cos \theta \, \pd/\pd r - r^{-1} \sin \theta \, \pd/\pd
\theta$ is the derivative along the axis of rotation. The
dots denote contributions from the Reynolds stress and molecular
viscosity \citep[e.g.][]{WKKB16}. The first term on the rhs describes
the effect of rotation, essentially the Coriolis force, on the mean
flow, whereas the second term corresponds to the baroclinic effect,
which results from latitudinal gradients of thermodynamic
quantities. In a perfect Taylor-Proudman balance the baroclinic term
vanishes and the isocontours of $\mean{\Omega}$ are cylindrical,
corresponding to $\pd\mean{\Omega}/\pd z =0$.

\begin{figure}
\begin{center}
    \includegraphics[width=.5\textwidth]{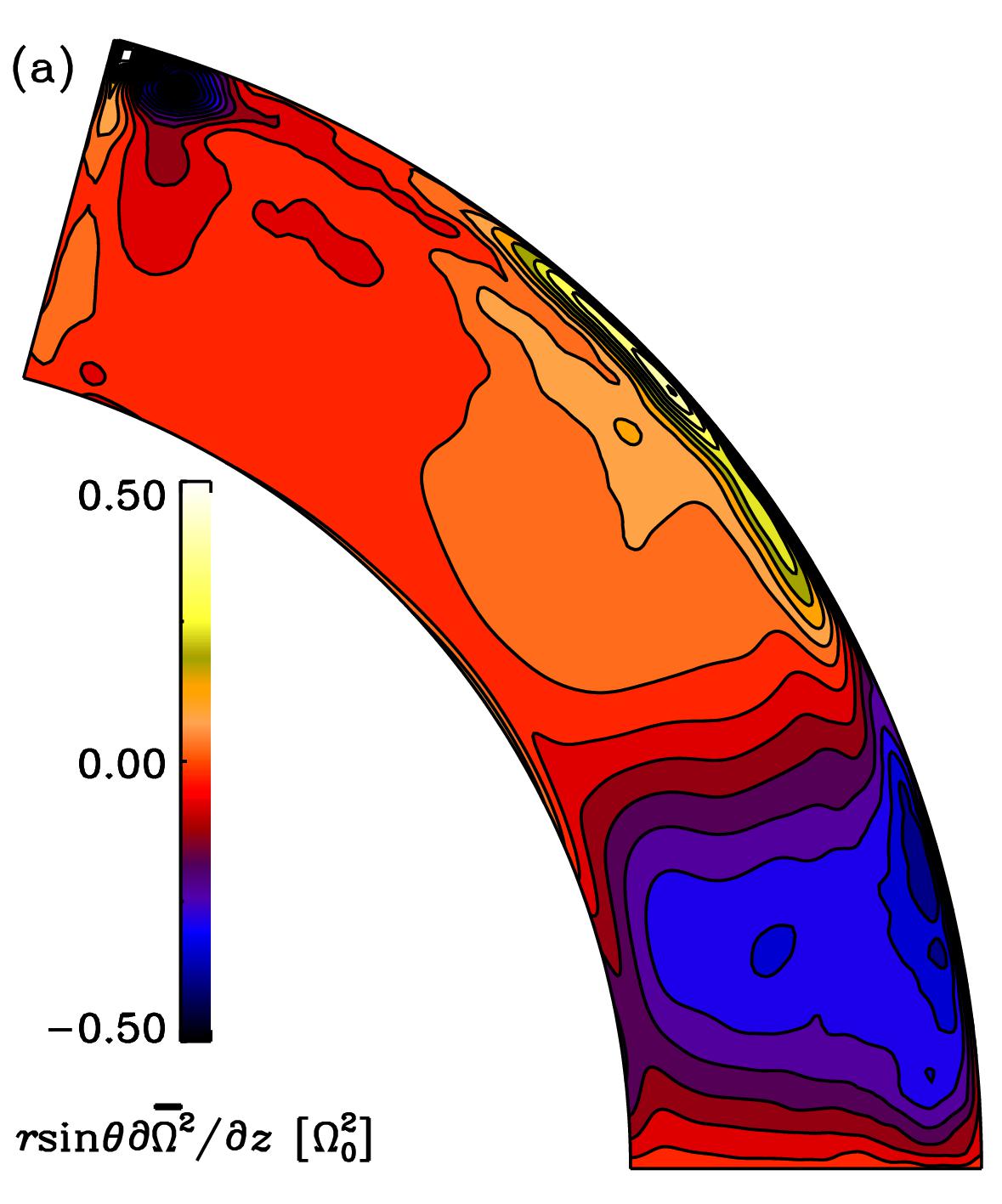}\includegraphics[width=.5\textwidth]{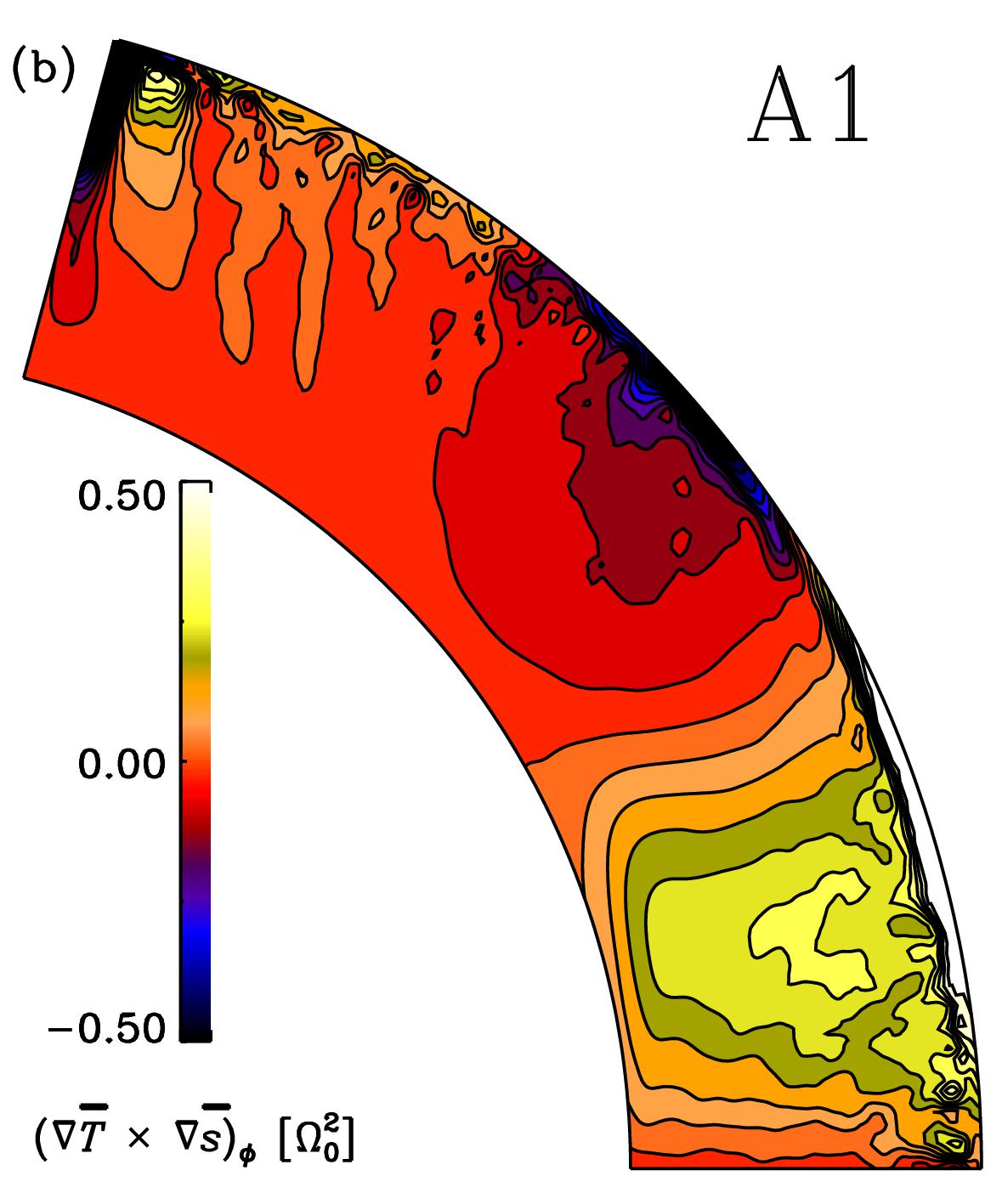}
    \caption{(a) Coriolis term from the mean vorticity equation
      (\ref{equ:vortphi}) from Run~A1 as a function of radius and
      latitude (b) The same as (a) but for the baroclinic term (colour online).}
\label{fig:baroclinic1}
\end{center}
\end{figure}

\begin{figure}
\begin{center}
    \includegraphics[width=\textwidth]{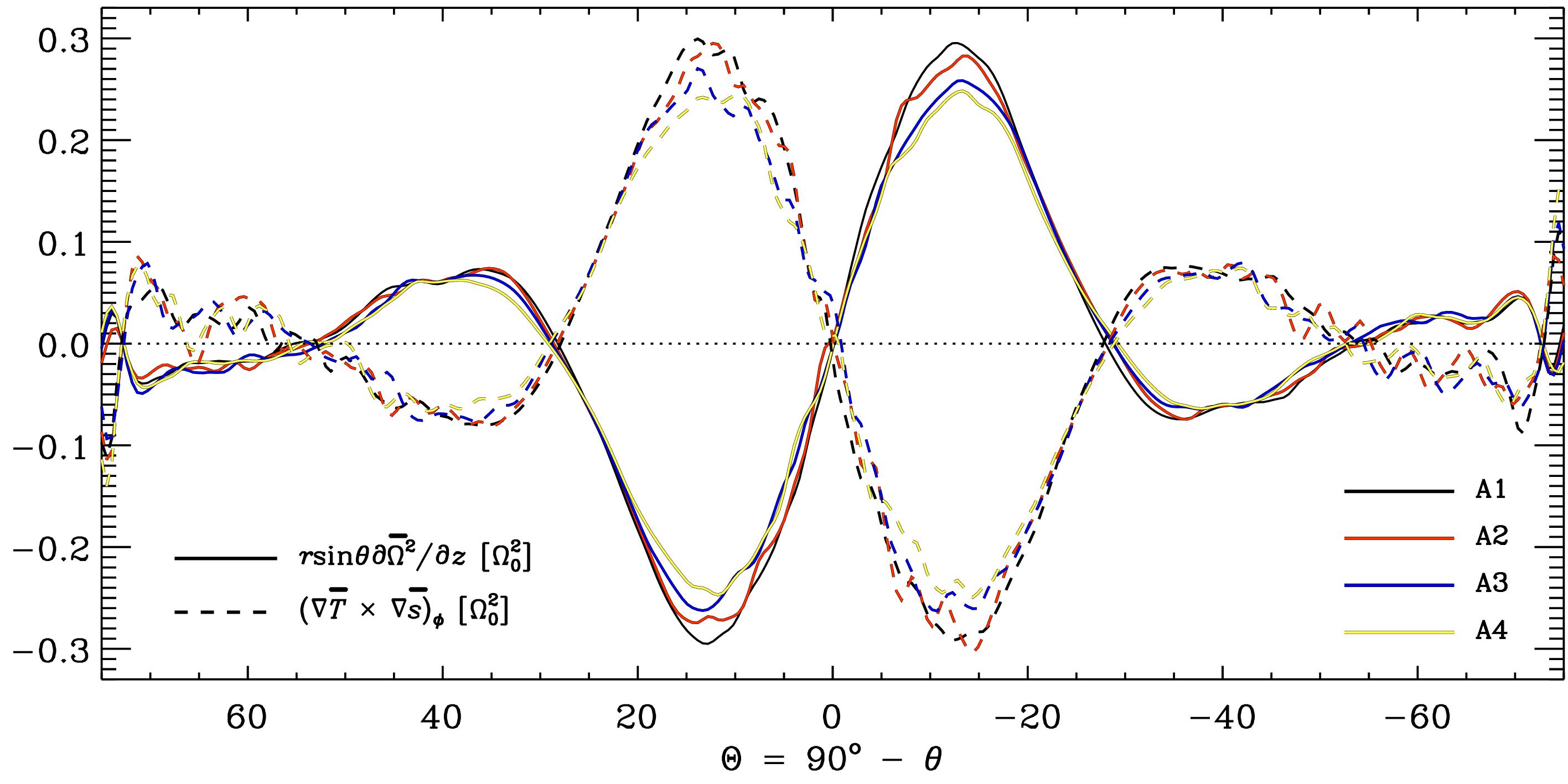}
    \caption{Coriolis (solid lines) and baroclinic (dashed) terms as
      functions of latitude $\Theta=90^\circ - \theta$ at the middle of the domain
      ($r=0.85R_\odot$) from Runs~A1 (black), A2 (red), A3 (blue), and
      A4 (yellow) (colour online).}
\label{fig:baroclinic2}
\end{center}
\end{figure}

Meridional cuts of the two terms on the right-hand side of (\ref{equ:vortphi}) from
Run~A1 are shown in \figu{fig:baroclinic1}. We find that the two terms
tend to balance in the bulk of the convection zone with larger
deviations occurring mostly near the surface. The current simulations
do not resolve the surface layers to a high enough degree to capture
the Reynolds stress-dominated region that is expected to occur there
\citep[e.g.][]{HRY15a}. \Figu{fig:baroclinic2} shows the Coriolis and
baroclinic terms as functions of latitude at the middle of the domain
$r=0.85R_\odot$ for Runs~A1--A4. In accordance with the similarity of
the rotation profiles, also the terms contributing to the baroclinic
balance are very similar in these runs; the only clear trend is a
slight decrease in the near-equator regions for both terms.
Thus, we conclude that the main effect of the decreasing luminosity is
a decrease in the Mach number, but this has only a weak
influence on the large-scale dynamics.

\subsection{Influence of the centrifugal force}

Typical stellar convection simulations either omit the contribution of the
centrifugal force or they consider it to be subsumed in
the gravitational force. This is also true for {\sc Pencil Code}
models, where the issue is more severe due to the enhanced rotation
rate. Here we study the influence of $\msFFF^{\rm cent}$ for the
first time in {\sc Pencil Code} simulations in spherical wedges.

We have introduced a parameter $c_{\rm cent}$ in front of the
centrifugal force in \equ{equ:centri}, with which it is possible to
regulate its strength. It is defined such that
\begin{eqnarray}
c_{\rm cent}\, = \,\bigl|\msFFF^{\rm cent}\bigr|\big/\bigl|\msFFF^{\rm cent}_0\bigr|\,,
\end{eqnarray}
where $\msFFF^{\rm cent}_0$ is the unaltered magnitude of the
centrifugal force. Such a procedure is used because the actual force
in the
simulations would be much stronger than in the Sun, for example. This is due
to the enhanced luminosity and rotation rate. Furthermore, the initial
condition is spherically symmetric and does not take the centrifugal
potential into account. Such a combination would lead to a violent readjustment in
the early stage of the simulation if $c_{\rm cent}=1$ was used.

\begin{figure}
\begin{center}
\begin{minipage}{150mm}
\subfigure[]{
\resizebox*{7cm}{!}{\includegraphics{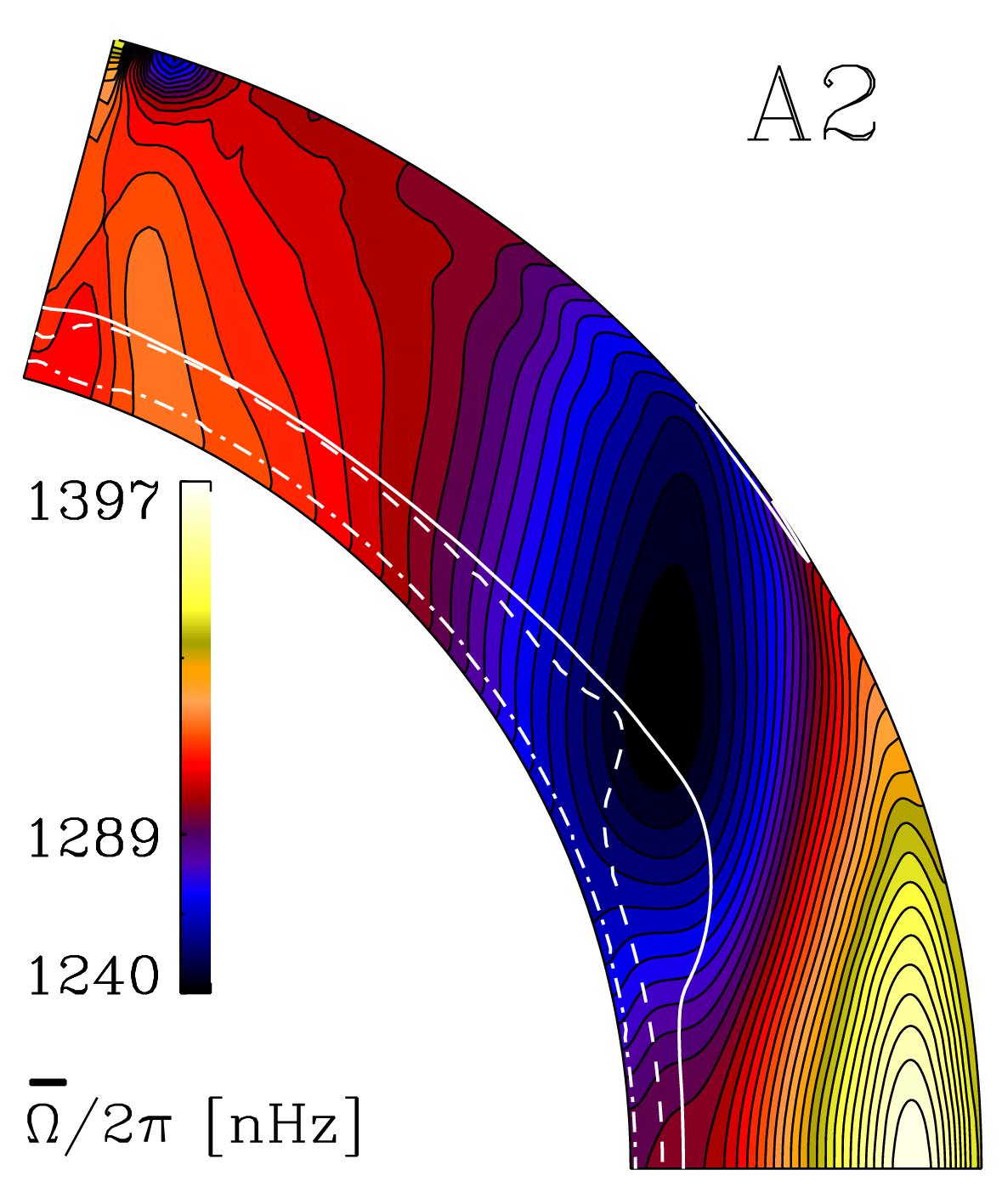}}}%
\subfigure[]{
\resizebox*{8cm}{!}{\includegraphics{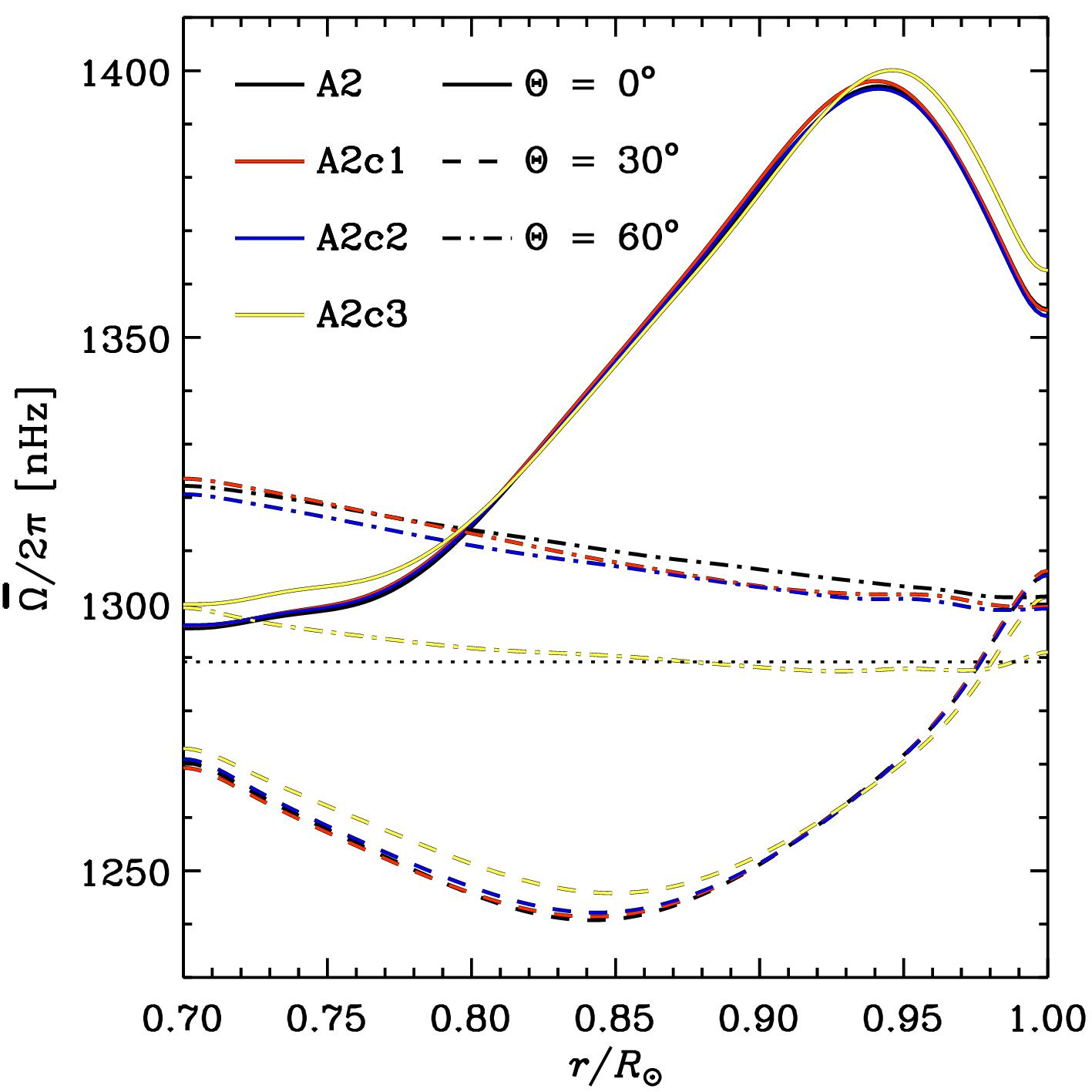}}}%
\caption{Same as \figu{fig:pOm_lumi} but for Runs~A2 (black), A2c1
  (red), A2c2 (blue), and A2c3 (yellow) (colour online).}
\label{fig:pOm_cent}
\end{minipage}
\end{center}
\end{figure}

We consider three cases where $c_{\rm cent}$ obtain values
$5\cdot10^{-4}$, $5\cdot10^{-3}$, and $0.05$ (Runs~A2c1, A2c2, and
A2c3 in \Tablel{tab:runs}) and compare those to a run with $c_{\rm
  cent}=0$ (Run~A2). Considering the ratio of the centrifugal force
and the acceleration due to gravity at the stellar surface at the
equator, these values translate to
\begin{eqnarray}
\bigl|\msFFF^{\rm cent}\bigr|\big/\bigl|\msFFF^{\rm grav}\bigr| \,\approx\, 2\cdot10^{-4} \ldots 0.02\,.
\end{eqnarray}
These are to be compared with the corresponding solar value,
\begin{eqnarray}
\bigl|\msFFF^{\rm cent}_\odot\bigr|\big/\bigl|\msFFF^{\rm grav}_\odot\bigr|\, = \,\Omega^2_\odot R_\odot/g_\odot \approx 2\cdot 10^{-5}\,.
\end{eqnarray}
Thus even the lowest value of $c_{\rm cent}$ considered here
corresponds to a relative strength of the centrifugal force that is an
order of magnitude greater than in the Sun.

In \figu{fig:pOm_cent} we compare the rotation profiles of the runs
where $c_{\rm cent}\neq0$ with that of Run~A2. We find that the
differences are minor with the exception of the high latitudes
($\Theta=60^\circ$) for Run~A2c3. The effect is relatively minor even
in this case, and considering that the magnitude of the
centrifugal force is already three orders of magnitude greater than in
the Sun, we estimate that its effect is likely to be minor in real
stars.
We note, however, that the cooling applied in the current simulations
is spherically symmetric and it is likely to work against the
centrifugal force.

\begin{figure}
\begin{center}
    \includegraphics[width=.6\textwidth]{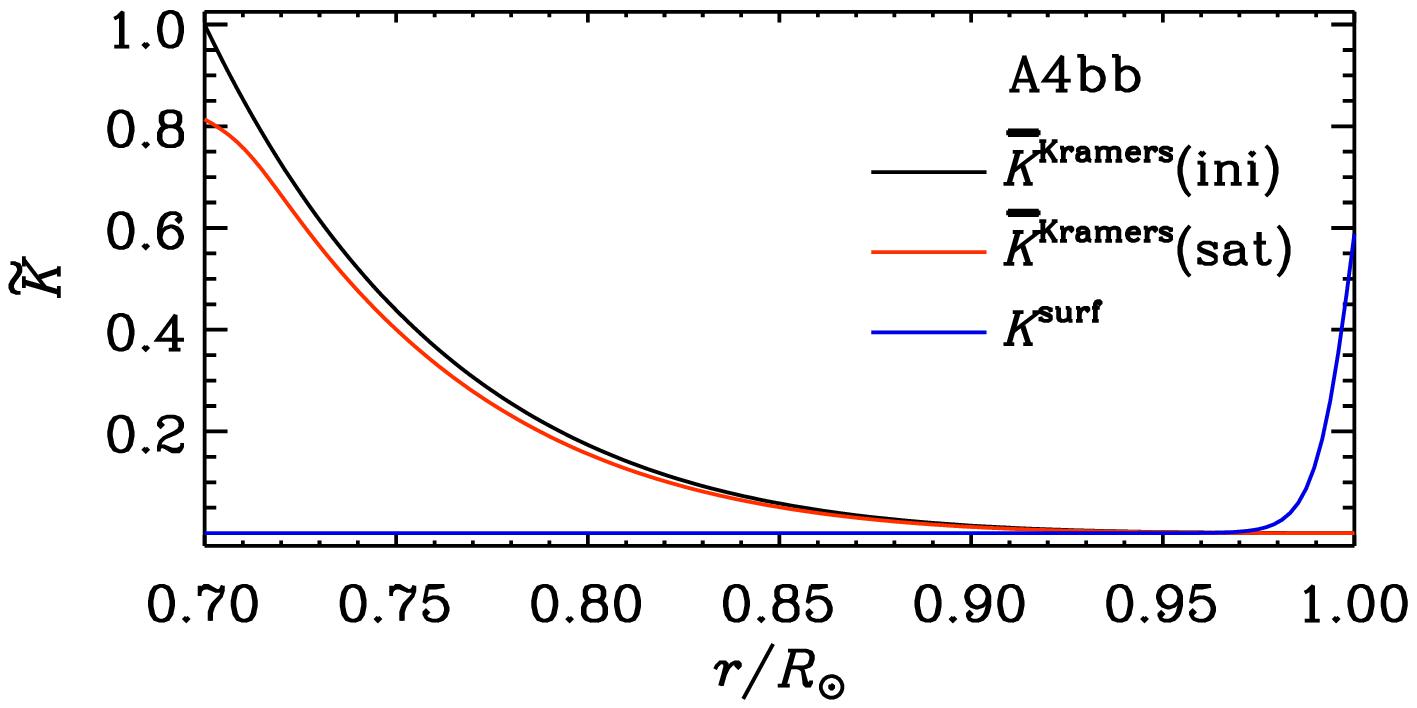}
    \caption{Initial (black) and saturated (red) profiles of
      $\mean{K}^{\rm Kramers}$ and $K^{\rm surf}$ (blue) from
      Run~A4bb (colour online).}
\label{fig:pkappa}
\end{center}
\end{figure}

\subsection{Influence of thermal BCs}
\label{sec:thermalBCs}

Various thermal BCs and treatments of the unresolved
photosphere have been used in the literature. For example, the ASH
simulations often apply a constant entropy gradient
\citep[][]{BMT04,BBBMT08} or a constant value of specific entropy at the
surface \citep[][]{2018ApJ...859..117N}. Furthermore, the energy flux
is carried through the upper surface via SGS entropy diffusion
\citep[e.g.][]{ABBMT12}. Similar conditions are used also by
\cite{FF14}, whereas \cite{HRY14} and their following work assume zero
radial gradient of the entropy. Several other anelastic simulations
assume a constant entropy on both radial boundaries
\citep[e.g.][]{GDW12,2015ApJ...810...80S}. Another approach is to
apply a constant temperature \citep[][]{KKBMT10,MMK15} or a black body
condition \citep[e.g.][]{KMB11}, where the former is typically
associated with a cooling applied near the surface. In the latter, the
flux at the surface is carried again by SGS diffusion.

\begin{figure}
\begin{center}
\begin{minipage}{150mm}
\subfigure[A4]{
\resizebox*{7.5cm}{!}{\includegraphics{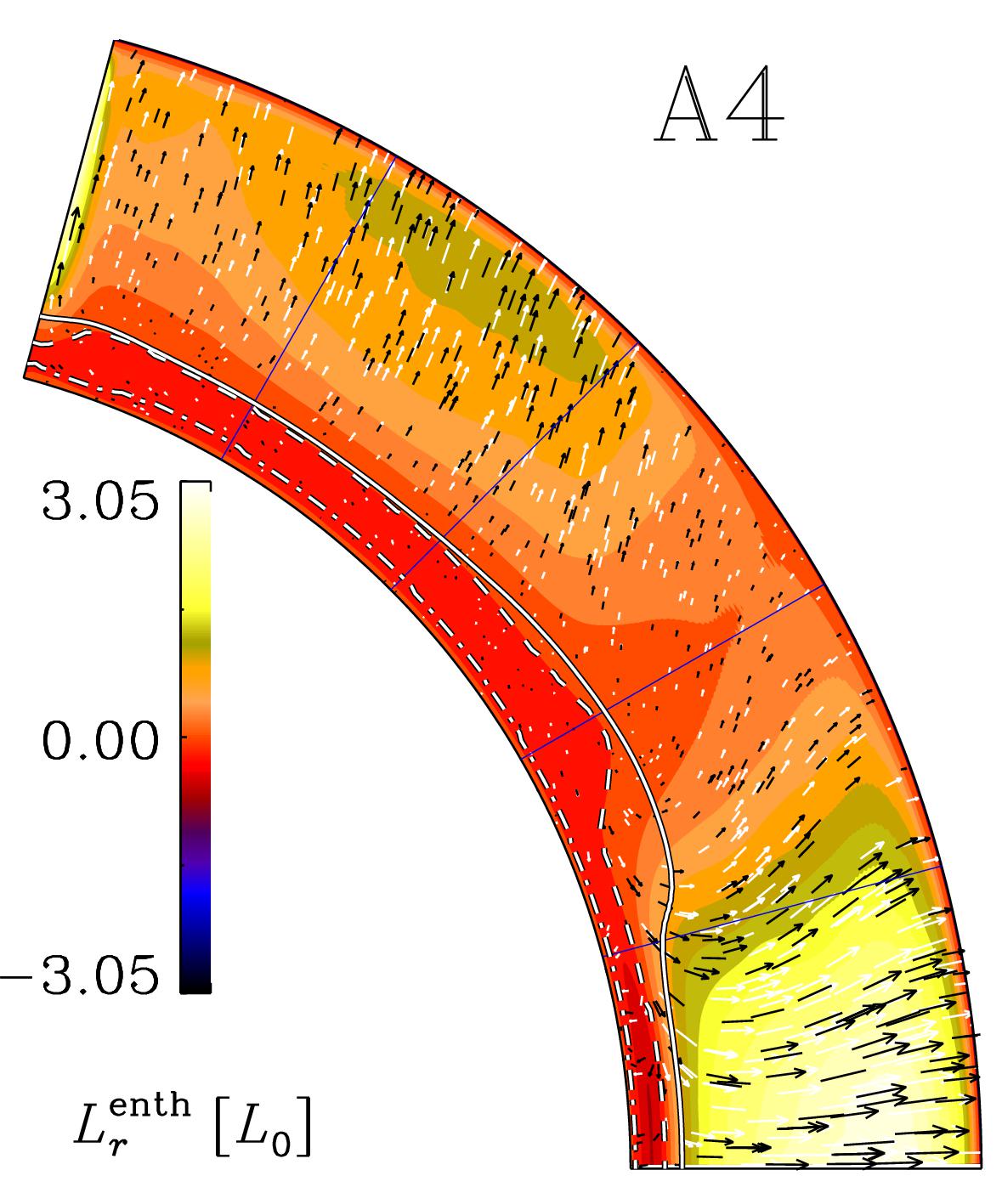}}}%
\subfigure[A4bb]{
\resizebox*{7.5cm}{!}{\includegraphics{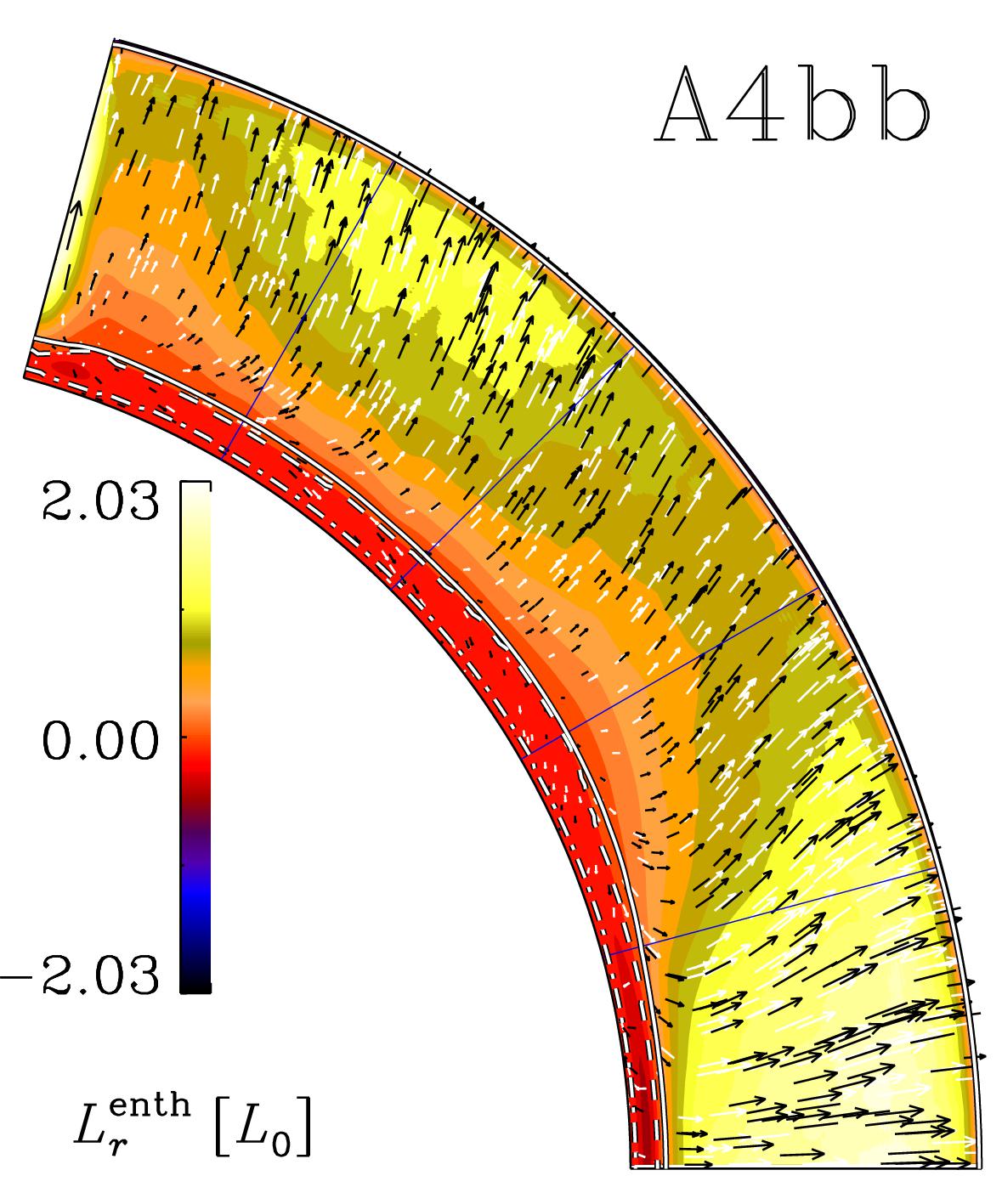}}}%
\caption{Radial enthalpy flux (colours) and the vectorial enthalpy
  flux (arrows) from Runs~A4 and A4bb. The solid, dashed, and
  dot-dashed black and white lines indicate the bottoms of the BZ, DZ,
  and OZ, respectively (colour online).}
\label{fig:pFent_ther}
\end{minipage}
\end{center}
\end{figure}

We consider two main setups where we either apply cooling in a shallow
layer with a constant temperature [cT, Eq.~(\ref{equ:cT})] imposed at
the surface (Run~A4) or enhanced radiative heat conductivity $K^{\rm
  surf}$ near the
surface (see \figu{fig:pkappa}) in conjunction with a black body [bb,
  Eq.~(\ref{equ:bb})]
condition (Run~A4bb). Both runs were repeated with a vanishing
entropy gradient at the surface (Runs~A4ds and A4ds2, respectively).

\begin{figure}
\begin{center}
    \includegraphics[width=\textwidth]{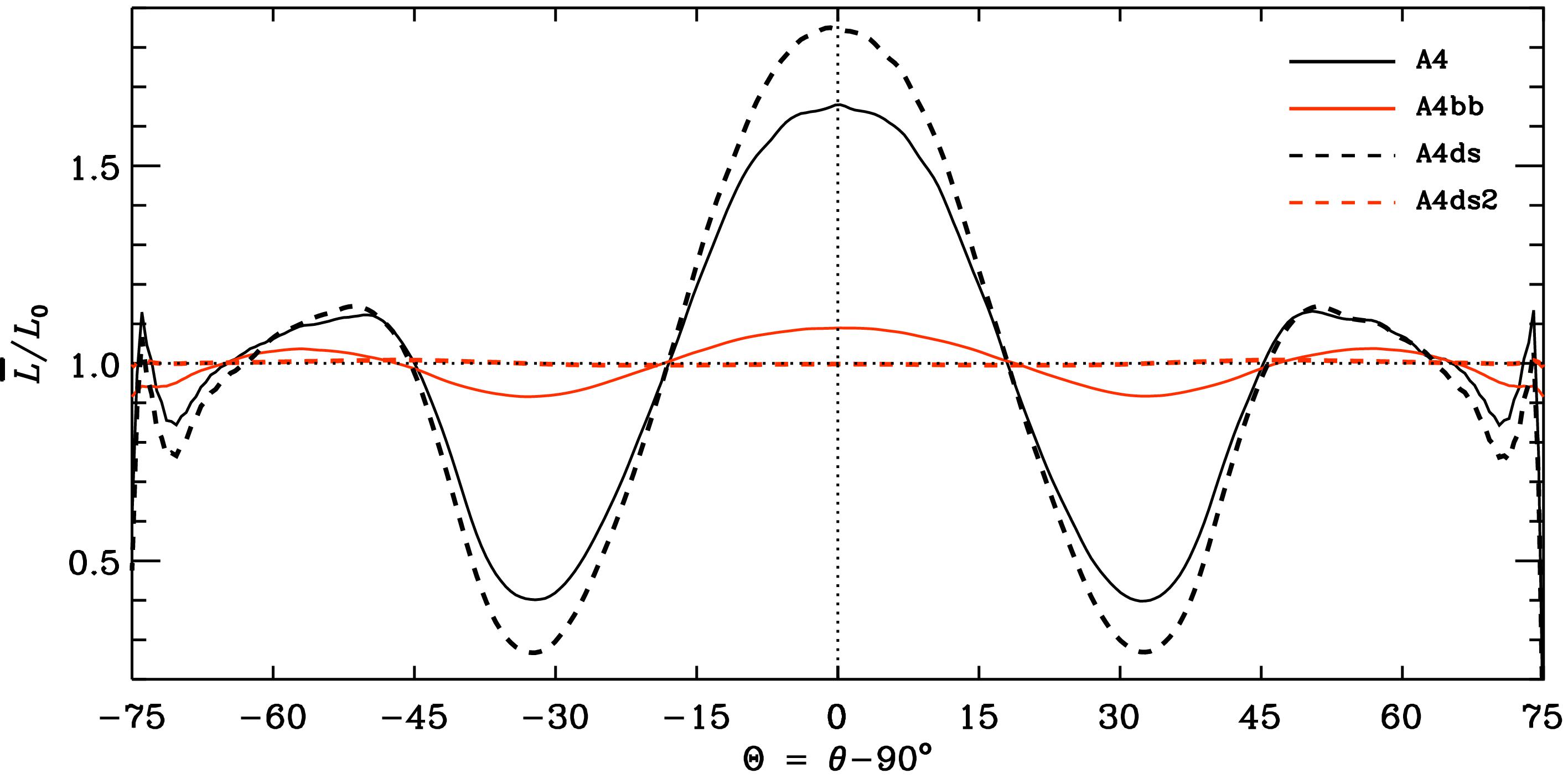}
    \caption{The total time averaged luminosity at $r=R_\odot$ from
      Runs~A4 (black solid line) and A4bb (red solid), A4ds (black
      dashed), and A4ds2 (red dashed) (colour online).}
\label{fig:psurf_flux}
\end{center}
\end{figure}

The convective energy transport, quantified by the luminosity of the
radial enthalpy flux $L_r^{\rm enth} = 4\pi r^2 \Fenthr$, is highly
anisotropic in Run~A4 with
the surface cooling and constant temperature BC;
see \figu{fig:pFent_ther}(a). Furthermore, the latitudinal
variation of the depth of the
buoyancy, overshoot, and Deardorff zones is substantial. We also note
the very weak convection around $\Theta=30^\circ$. An earlier study
\citep{2018arXiv180305898K} has shown that in an otherwise identical
setup, but where a fixed profile of $K$ is used, leads to a situation
where only a very thin surface layer is convectively unstable
(e.g.\ their Run~MHDp). In
Run~A4bb, the black body
condition is used in addition to enhanced radiative diffusion near the
surface, transporting the energy through the surface. In this case the
convective energy transport is clearly less anisotropic than in
Run~A4, although substantial latitudinal variation still occurs; see
\figu{fig:pFent_ther}(b). Furthermore, \figu{fig:psurf_flux} shows
that the surface luminosity varies much more in Run~A4 than in
Run~A4bb.
The extreme latitude dependence in Run~A4 can be explained by the fact
that the flux near the surface is determined by the difference between
a fixed spherically symmetric profile of the temperature $T_{\rm
  cool}$ and the dynamically evolving
actual temperature $T$:
\begin{eqnarray}
F_r^{\rm cool}\, = \,\int_{r_0}^{R_\odot} \Gamma_{\rm cool} \, {\rm d}r\, = \,-\,\Gamma_0 \int_{r_0}^{R_\odot} f(r) (T_{\rm cool}-\brac{T}_{\theta \phi}) \, {\rm d}r\,.
\end{eqnarray}
Note that in the cases with surface cooling, the radiative flux at the
surface is negligible and $L^{\rm cool} = 4\pi r_1^2 F_r^{\rm cool}
\approx L_0$. At mid-latitudes, the actual temperature has
a local minimum, and the cooling due to the relaxation term in the
entropy equation becomes inefficient, as seen in
\figu{fig:psurf_flux}. This leads to a more stable thermal
stratification at mid-latitudes ($20\lesssim|\Theta|\lesssim45$). The
situation is qualitatively similar although the latitudinal variation
is even slightly enhanced in Run~A4ds where a vanishing
radial entropy gradient is enforced at the surface.

In the case of Run~A4bb, however, the flux is carried by radiative
diffusion near the surface, which is
proportional to the radial derivative of the temperature, which varies
much less as a function of latitude than the difference between a
fixed reference temperature and the actual value of $T$. There is
still substantial latitudinal variation, on the order of 10 per cent
of the total luminosity. This is due to the non-linear nature of the
black body BC, see \equ{equ:bb}:
\begin{eqnarray}
-K^{\rm tot} \frac{\pd T}{\pd r} \,= \,\sigma T^4\,,
\end{eqnarray}
where $K^{\rm tot} = K^{\rm Kramers} + K^{\rm surf}$. In practise
$K^{\rm Kramers}\ll K^{\rm surf}$ near the surface and $F_r^{\rm rad}\approx -
K^{\rm surf} {\pd T}/{\pd r}$. However, adopting the `ds' BC
(Run~A4ds2) leads, under the assumption of hydrostatic equilibrium, to
${\pd T}/{\pd r}=g/c_{\rm P}$ which is independent of latitude and
time. This implies that the radiative (=total) flux is fixed at both
boundaries which is indeed reproduced by the simulation, see the red
dashed line in \figu{fig:psurf_flux}. However, the total energy in this
simulation does not find a saturated state but a constant drift is
observed as a function of time. This is an issue related to having von
Neumann type BCs at both boundaries.
We find that the choice of thermal BC has a relatively minor effect on the
surface luminosity and that the results are more sensitive to the
parameterisation of the photospheric physics. The only exception is
the case where a constant radiative flux is imposed at both boundaries
(Run~A4ds2) which leads to an unphysical drift of the total energy of
the solution.

\begin{figure}
\begin{center}
\begin{minipage}{150mm}
\subfigure[A4]{
\resizebox*{7.5cm}{!}{\includegraphics{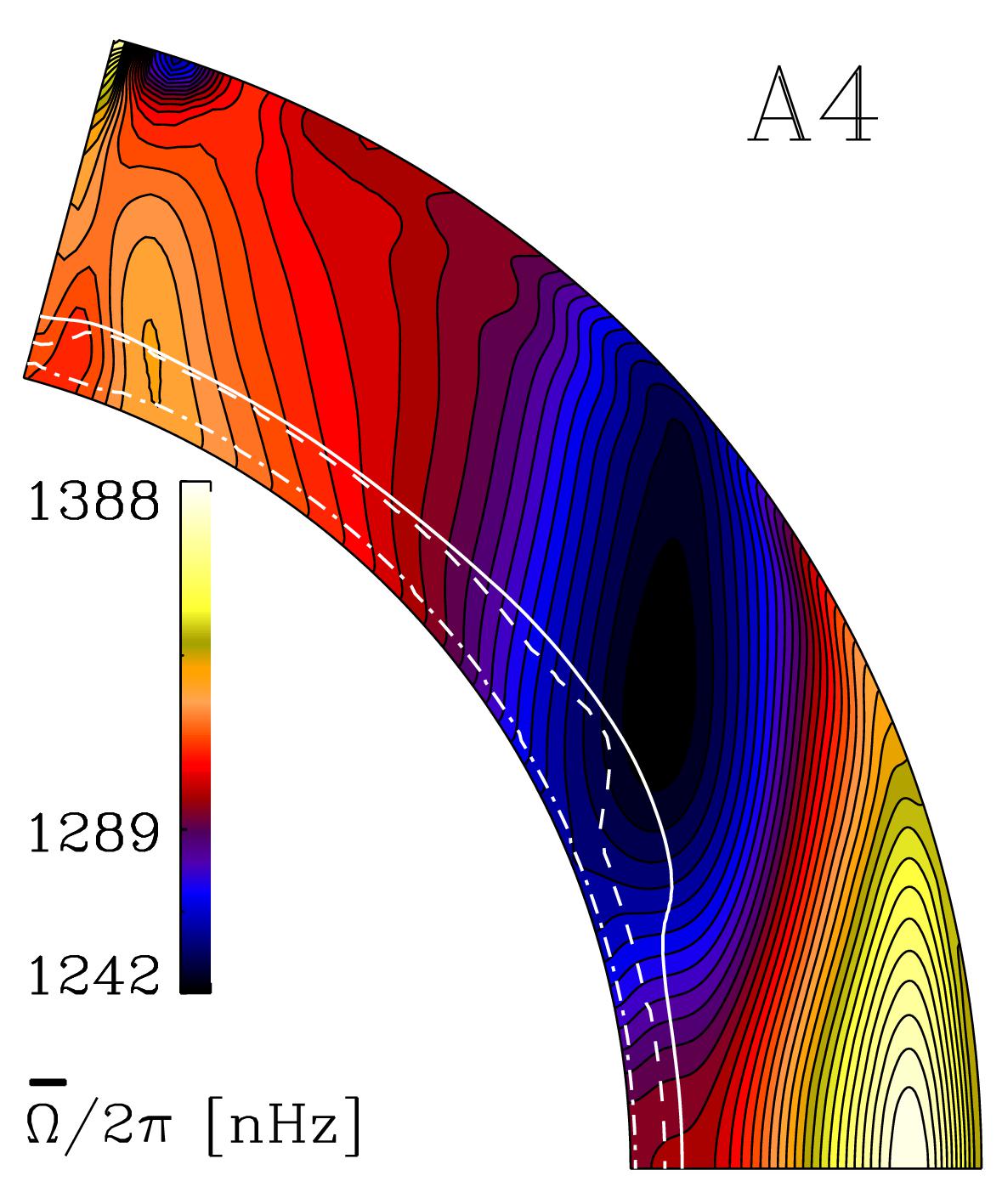}}}%
\subfigure[A4bb]{
\resizebox*{7.5cm}{!}{\includegraphics{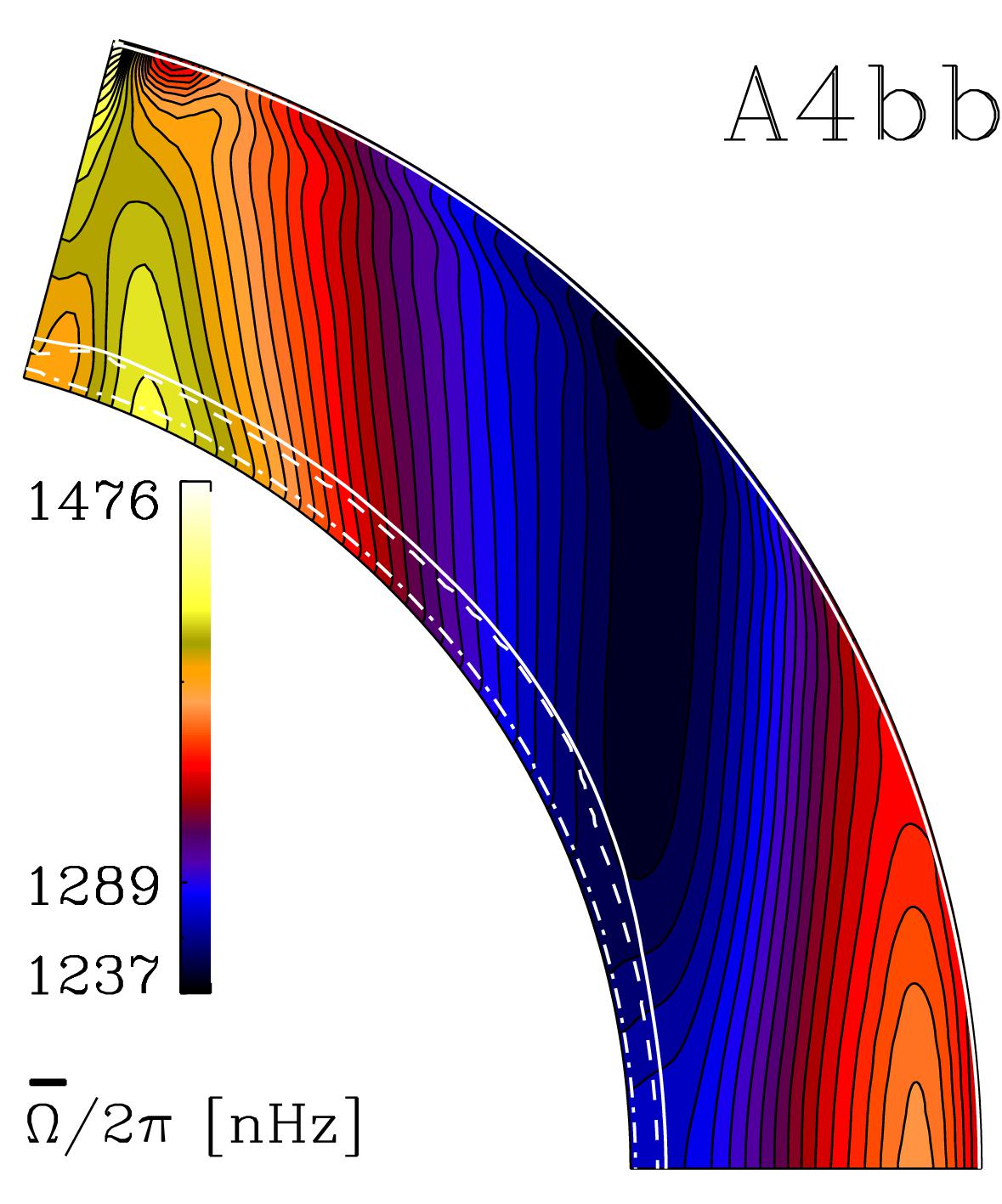}}}%
\\
\caption{Temporally averaged mean angular velocity
  $\mean{\Omega}=\mean{U}_\phi/r \sin\theta + \Omega_0$ from
  Runs~A4 and A4bb (colour online).}
\label{fig:pOm_ther}
\end{minipage}
\end{center}
\end{figure}

We find a substantial poleward contribution to the heat flux in all
rotating cases; see the arrows for $\mean{\bm{F}}^{\rm
  enth}=(\Fenthr,\Fentht,0)$ in \figu{fig:pFent_ther}. The tendency
for the enthalpy flux to align with the rotation vector is an
established result from mean-field theory of hydrodynamics
\citep{R89,KPR94}. Furthermore, mean-field models have shown that such
poleward flux is instrumental in producing a pole-equator temperature
difference that can break the Taylor-Proudman balance \citep{BMT92}.

The rotation profiles from Runs~A4 and A4bb are shown in
\figu{fig:pOm_ther}. We find that the cases with surface
cooling deviate more strongly from the Taylor-Proudman
balance. Furthermore, the latitudinal variation of the bottom of the
buoyancy and overshoot zones are more pronounced in these cases. The
runs with diffusive transport of thermal energy near the surface
also tend to exhibit strong polar vortices. However, this feature is
likely to be dependent on the initial conditions or the history of the
run, as was shown by \cite{GYMRW14} and \cite{KKB14}.
We again find that the choice of BC is less
important than the treatment of the photosphere. The rotation
profiles of Runs~A4 and A4ds are practically identical despite the
different boundary conditions. The averaged angular velocities in
Runs~A4bb and A4ds2 are also qualitatively similar, despite the fact
that the kinetic energy in the latter is slowly increasing.

\subsection{Influence of magnetic BCs}

Here we compare the dynamo solution of Run~M1 from \cite{KKOBWKP16} with the vE
magnetic BC with a corresponding Run~M2 with the vJ BC of
\cite{GKW17}.
While the vE conditions assume that the electric field vanishes,
they allow non-vanishing horizontal currents on the boundary. The vJ
conditions assume that also the currents vanish on the
boundary.
In spherical coordinates the tangential components of the current
density are given by
\begin{align*}
J_\theta \,=\,&\,
       \dfrac{1}{r^2\sin \theta}
       \dfrac{\upartial^2 A_\phi}{\upartial \theta\upartial \phi}
       +\dfrac{\cot\theta}{r^2\sin \theta}
       \dfrac{\upartial A_\phi     }{\upartial \phi}
       -\dfrac{1}{r^2\sin^2 \theta}
       \dfrac{\upartial^2 A_\theta}{\upartial \phi^2}\\
      &\hskip 23mm  -
       \dfrac{\upartial^2 A_\theta}{\upartial r^2}
       -\dfrac{2}{r}\dfrac{\upartial A_\theta}{\upartial r}
       +\dfrac{1}{r}\dfrac{\upartial^2 A_r     }{\upartial r\upartial \theta}\,,\\
J_\phi\, =\,&\,
       \dfrac{1}{r^2\sin \theta}
       \left(
       \dfrac{A_\phi}{\sin \theta}
       -\cos\theta\dfrac{\upartial  A_\phi  }{\upartial \theta}
       -\cot\theta\dfrac{\upartial A_\theta}{\upartial \phi}
       +\dfrac{\upartial^2 A_\theta}{\upartial \theta\upartial \phi}
       +r\dfrac{\upartial^2 A_r }{\upartial r\upartial \phi}
       \right)\\
       &\hskip 23mm -\dfrac{\upartial^2 A_\phi}{\upartial r^2}
       - \dfrac{2}{r}\dfrac{\upartial A_\phi}{\upartial r}
       -\dfrac{1}{r^2}\dfrac{\upartial^2 A_\phi}{\upartial \theta^2}\,.
\end{align*}
The terms involving $A_r$ vanish on the boundary under the condition
${\upartial A_r }/{\upartial r}=0$.
Setting $A_\theta$ and $A_\phi$ constant on the boundary (e.g., 0),
eliminates the remaining terms involving the tangential derivatives,
see \equ{equ:vE}.
There remains an additional constraint for the horizontal components
of $\AAA$ satisfying
\begin{equation}
       \dfrac{\upartial^2 A_\theta}{\upartial r^2}
       + \dfrac{2}{r}\dfrac{\upartial A_\theta}{\upartial r}\, =\, 0\,,\hskip 20mm
       \dfrac{\upartial^2 A_\phi}{\upartial r^2}
       + \dfrac{2}{r}\dfrac{\upartial A_\phi}{\upartial r}\, =\, 0\,.
\end{equation}
We recognize that this is technically over-determined, with five BCs on three
equations, and a more general solution to the BC would be
desirable.

Apart from the BCs, the models differ through the inclusion of a set of test
fields \citep[see e.g.,][]{SRSRC05,SRSRC07,2018A&A...609A..51W}.
These are used to extract numerically the turbulent transport coefficients
responsible for the evolution of large-scale magnetic fields in the
framework of mean-field dynamo theory \citep[e.g.][]{M78,KR80}.
The test fields are acted upon by the flow, generated by the MHD solution,
but, unlike the physical magnetic field, there can be no feedback on the flow nor
on the energy via Lorentz force and Ohmic heating, respectively.
The solution should therefore be independent of the test fields.
However, the Courant condition is also applicable to the evolution of the test
fields and typically necessitates a slightly reduced time step.
Due to the chaotic nature of such a system, the details of the
solutions diverge, but the statistical properties such as cycle
lengths remain consistent.

To examine the potential differences in the solutions accounted for by the
BCs, we consider equally long and similar epochs in the dynamo
solutions for both models. The chosen epoch represents a solar-like
state of the solutions. Such states occur at different times in the
two simulations due to the changes in the length of the time step.
In this context we mean by `solar-like' that near the surface the azimuthal
magnetic field exhibits a regular equatorward drift in lower latitudes
and poleward drift in higher latitudes.
The magnetic field shows cyclic polarity reversals and typically has
opposite signs on the two hemispheres (antisymmetric with respect to the
equator).
As has been described in detail in \cite{KKOBWKP16}, such regular
epochs are rather rare in these simulations, as especially the parity can undergo changes
to nearly symmetric solutions (i.e., the same orientation of the toroidal field
in both hemispheres), the migration patterns, however, remaining unaltered.

\begin{figure}[t!]
\begin{center}
\begin{minipage}{150mm}
\begin{center}
{\resizebox*{12.0cm}{!}{\includegraphics[trim=0.3cm 0.45cm 1.0cm  0.3cm,clip=true]{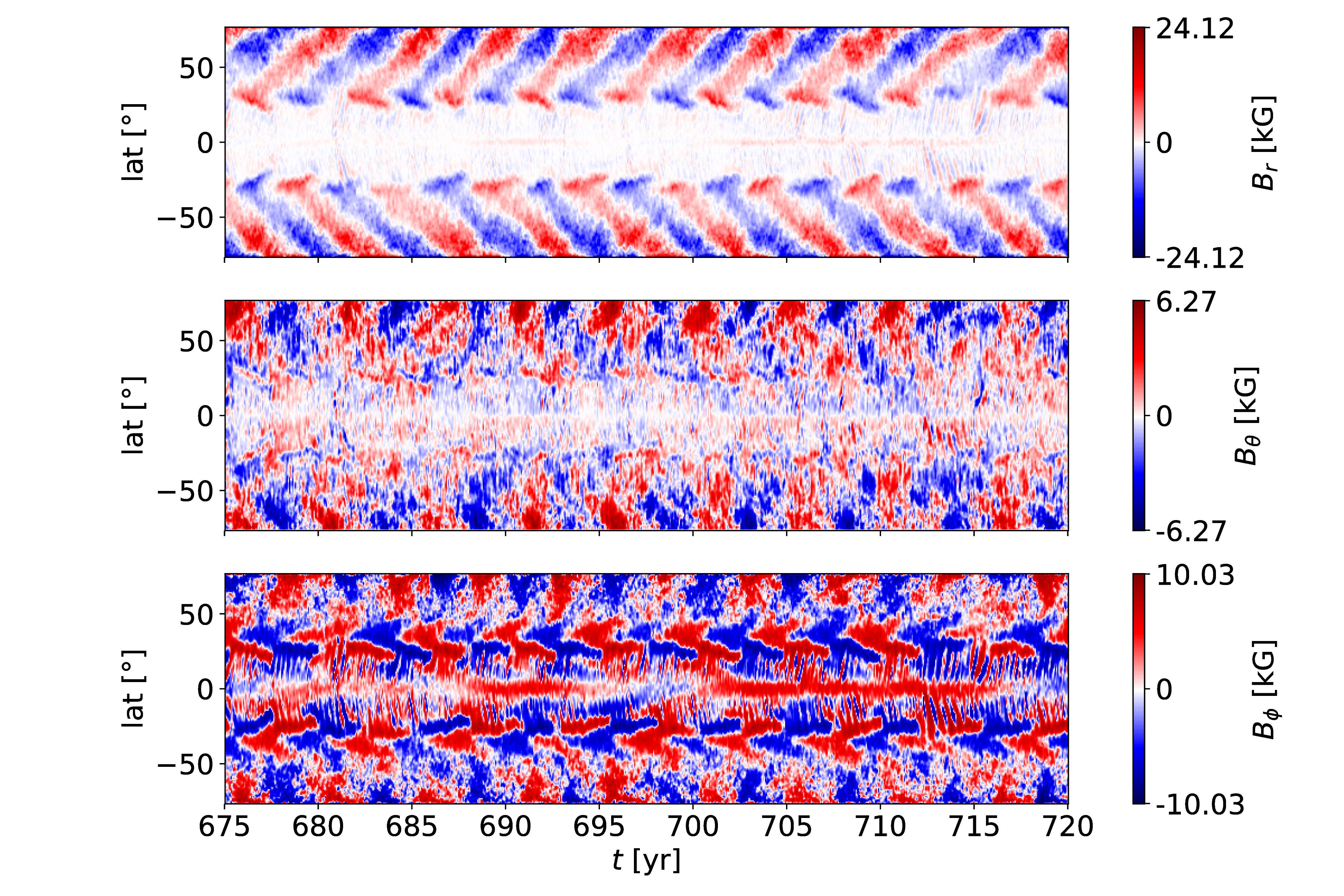}}}\\%
{\resizebox*{12.0cm}{!}{\includegraphics[trim=0.3cm 0.45cm 1.0cm -0.5cm,clip=true]{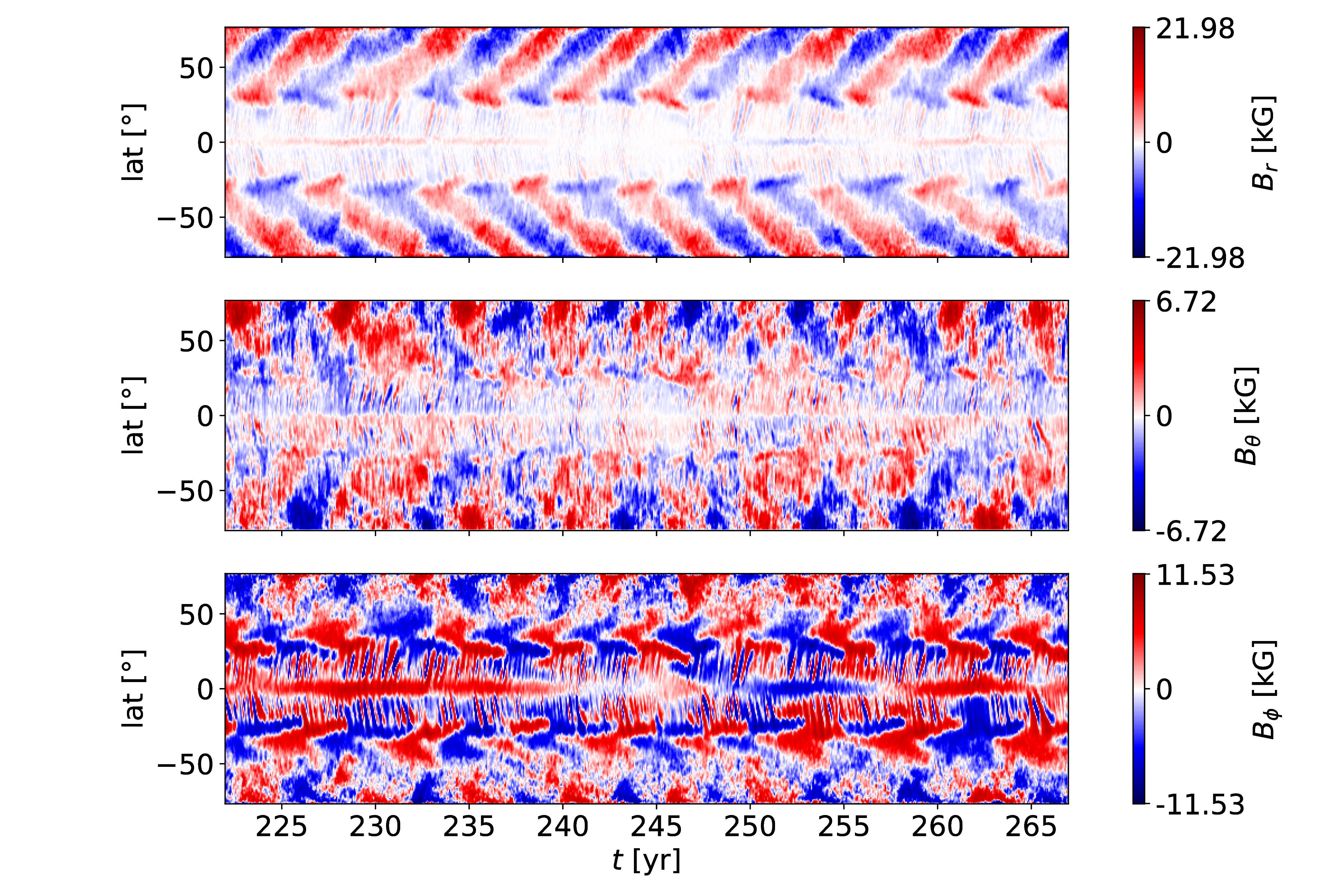}}}%
\caption{
Near-surface ($r=0.98R_\odot$) magnetic field
butterfly diagrams from
Runs~M1
(top)
and M2
(bottom) (colour online).
}%
\label{fig:Bsurf}
\end{center}
\end{minipage}
\end{center}
\end{figure}

\begin{figure}[t!]
\begin{center}
\begin{minipage}{150mm}
\begin{center}
{\resizebox*{12.0cm}{!}{\includegraphics[trim=0.3cm 0.45cm 1.0cm  0.3cm,clip=true]{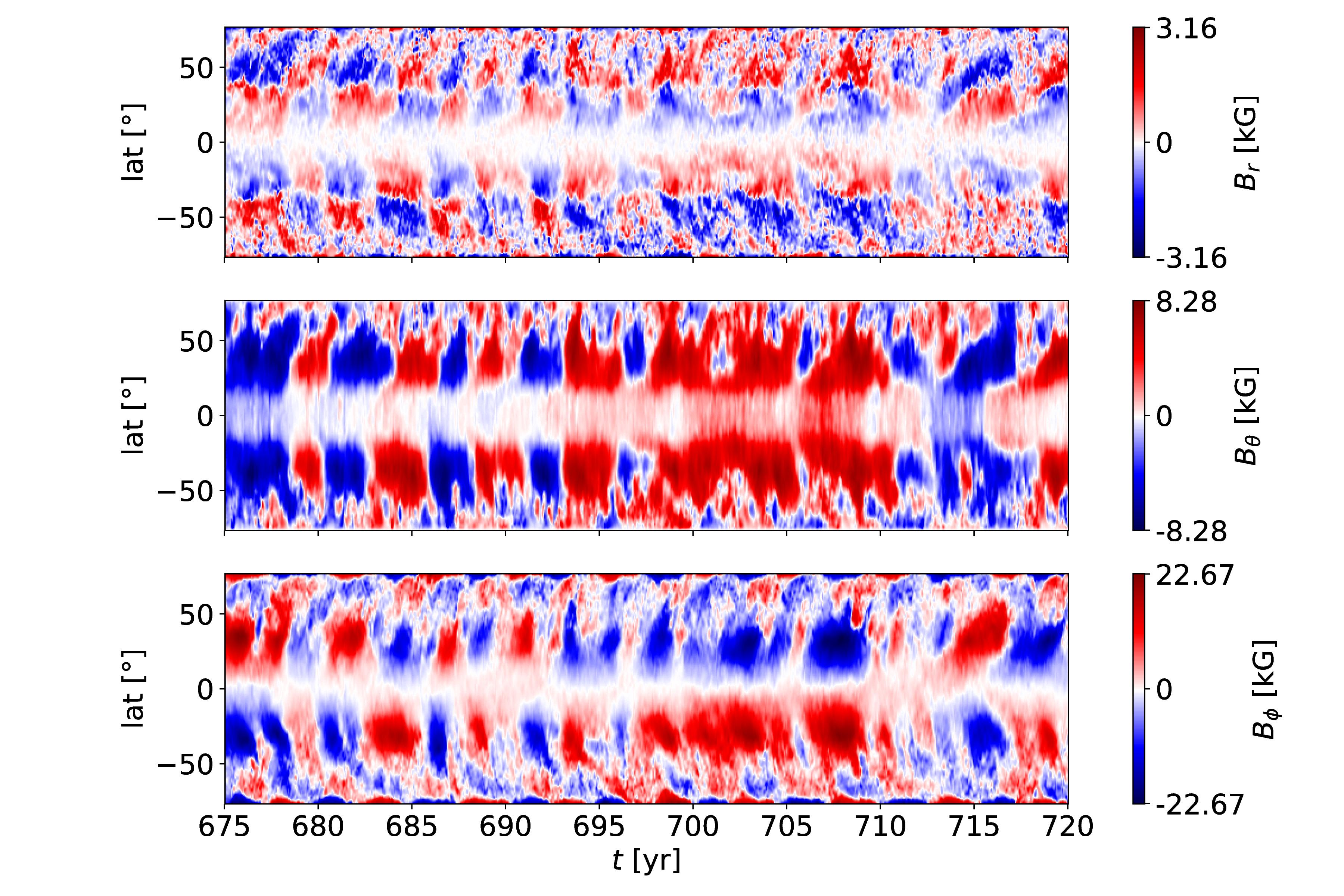}}}\\%
{\resizebox*{12.0cm}{!}{\includegraphics[trim=0.3cm 0.45cm 1.0cm -0.5cm,clip=true]{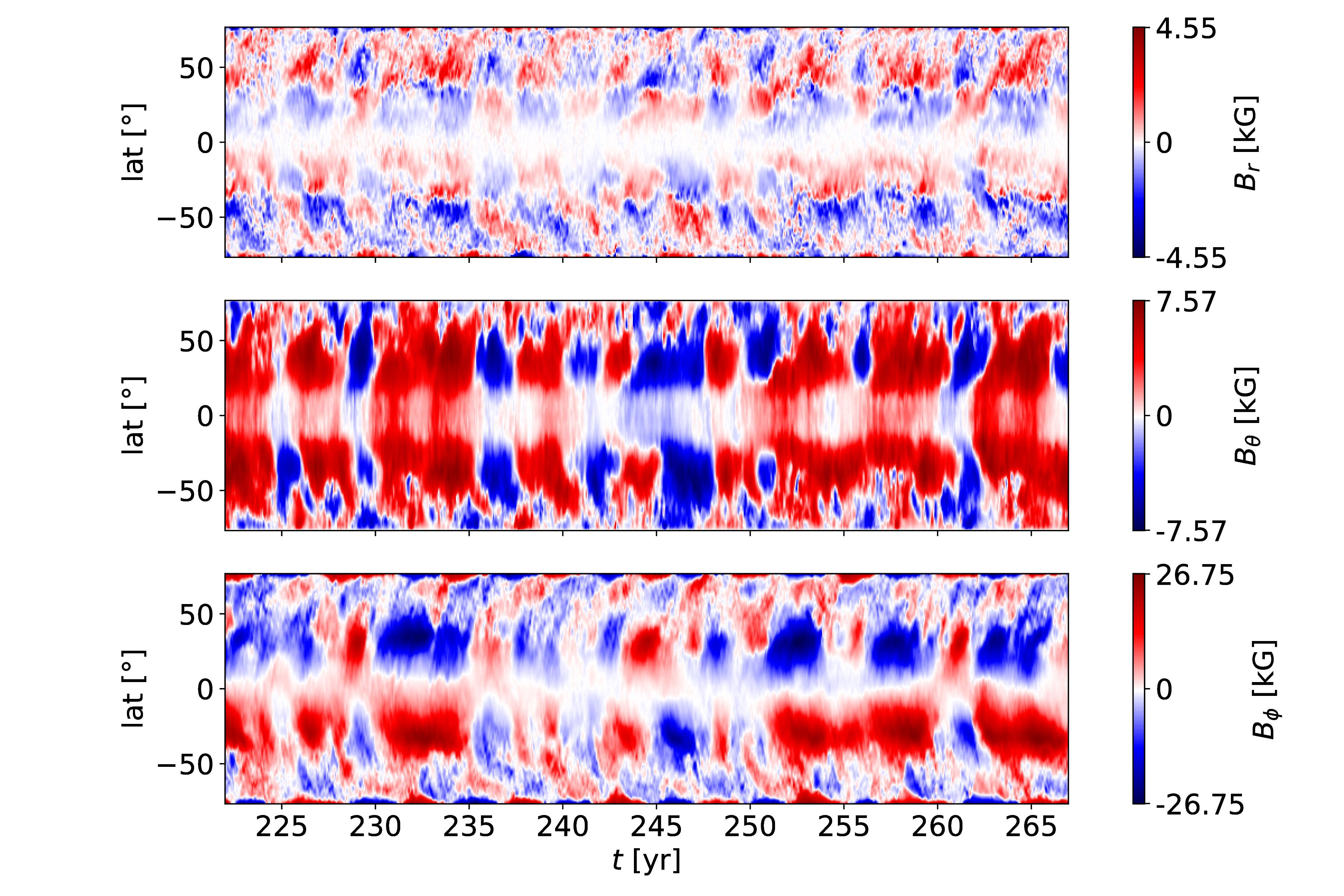}}}%
\caption{
Base ($r=0.72R_\odot$) magnetic field butterfly diagrams from
Runs~M1 (top)
and M2
(bottom) (colour online).
}%
\label{fig:Bbase}
\end{center}
\end{minipage}
\end{center}
\end{figure}

\Figu{fig:Bsurf} depicts the solar-like solution near the surface
of the convection zone, $r=0.98 R_\odot$
by magnetic field component from each of Runs~M1 with vE BCs
(upper three panels) and M2 with vJ BCs (lower three panels).
As is evident from \figu{fig:Bsurf}, the runs with different boundary conditions do not
differ much. Also, the cycle period in Run~M2 appears slightly longer than in M1, while
the amplitude of the magnetic field is nearly unaffected.

We might expect the differences in the boundary conditions to be most apparent
near the base of the convection zone, hence in \figu{fig:Bbase} we show
time-latitude
diagrams close to the boundary in each Run~M1 and M2 at $r=0.72 R_\odot$.
There, we see two different incarnations of the long-period, nearly purely antisymmetric,
dynamo cycle described in detail by \cite{KKOBWKP16}.
Hence, the effect of the BCs on the overall dynamo solution are very
small, and part of the variation seen here
is also likely to arise from the intrinsically chaotic nature of the solutions.

\begin{figure}[t!]
\begin{center}
\begin{minipage}{150mm}
\begin{center}
{\resizebox*{12.0cm}{!}{\includegraphics[trim=1.3cm 0.2cm 1.8cm 2.0cm,clip=true]{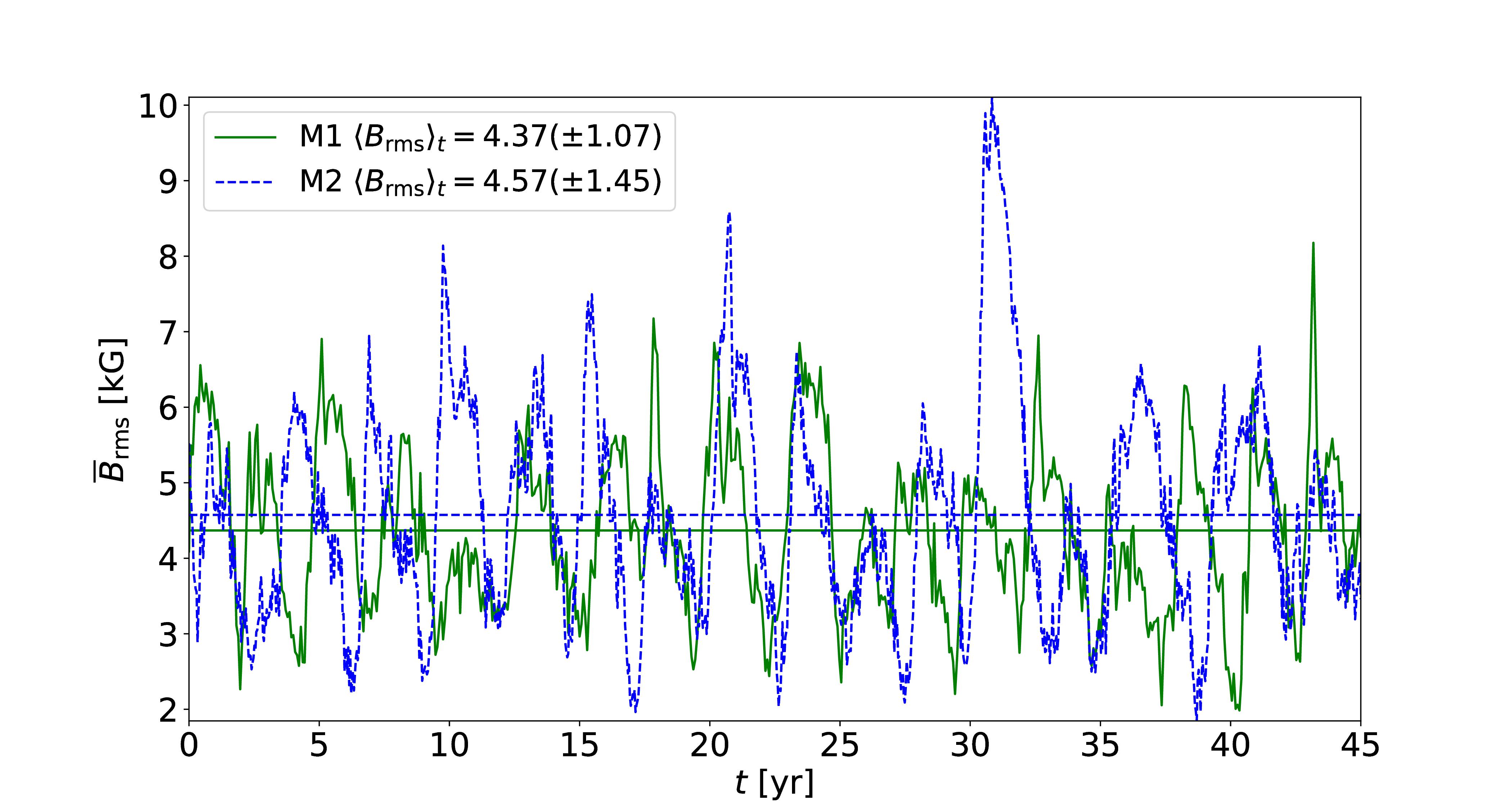}}}\\%
\caption{
Comparison of azimuthally averaged
magnetic field (rms) near the base of the convection zone
($r<0.73R_\odot$)
between Runs~M1
(green, solid) and M2
(blue, dashed) during the
respective 45-year time intervals 222--267 and 675--720 years (colour online).
}%
\label{fig:brms}
\end{center}
\end{minipage}
\end{center}
\end{figure}

As an additional check on the impact of the BCs on the solution, we also
compare the evolution of the rms of the azimuthally averaged magnetic field
strength in Runs~M1 and M2 during this 45 year period near the boundary.
The layer $r<0.73R_\odot$ is considered and the time evolution plotted in
\figu{fig:brms}.
The common time is initialised to zero for the purposes of the plot.
The temporal averages for $B_{\rm rms}$, during this period were computed as
4.37~kG and 4.57~kG with standard deviation of 1.07~kG and 1.45~kG for M1 and M2,
respectively.
This is a rather small difference, as we already concluded from the time-latitude
diagrams.

\begin{figure}[t!]
\begin{center}
\begin{minipage}{150mm}
\begin{center}
{\resizebox*{15.0cm}{!}{\includegraphics[trim=0cm 0.8cm 0cm  0cm,clip=true]{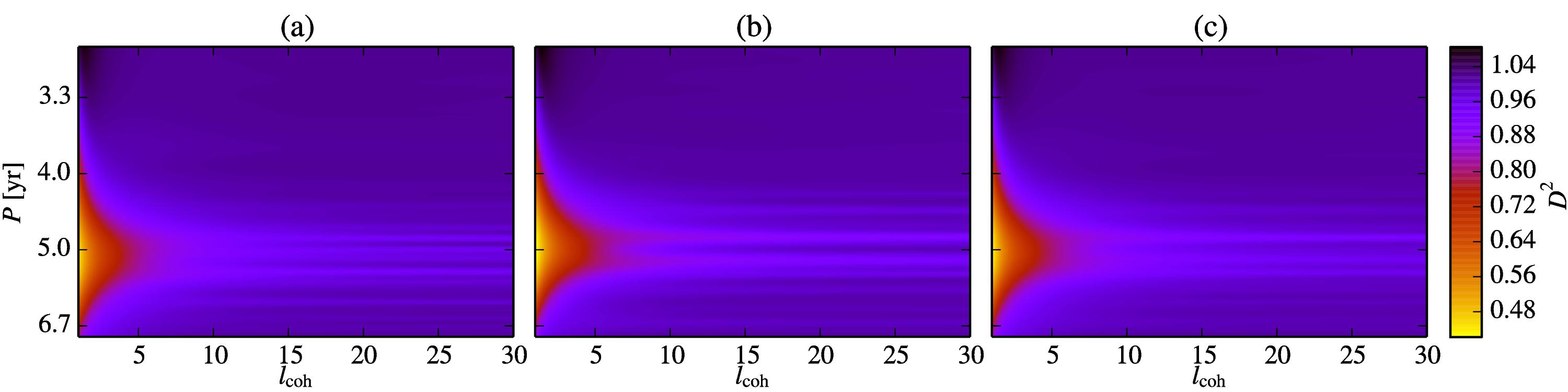}}}\\%
{\resizebox*{15.0cm}{!}{\includegraphics[trim=0cm 0.1cm 0cm 0.7cm,clip=true]{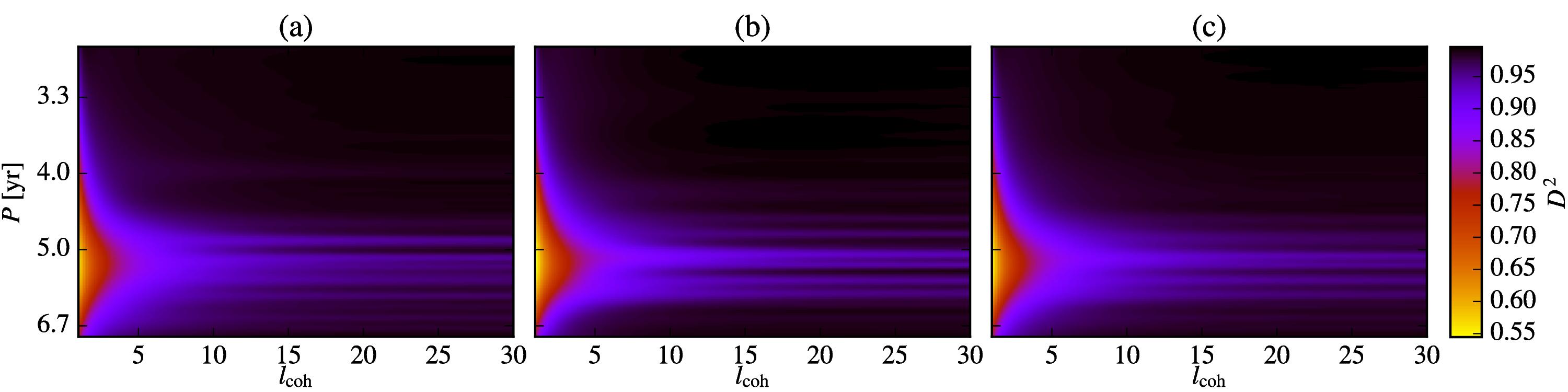}}}%
\caption{
Comparison of the $D^2$ spectra of azimuthally averaged $B_\phi$ for Runs~M1 (top)
and M2 (bottom).
Panel (a) corresponds to north, (b) to south, and (c) to full latitudinal extent (colour online).
}%
\label{fig:D2}
\end{center}
\end{minipage}
\end{center}
\end{figure}

To reveal the differences in more detail, we repeat the analysis
used to determine the basic dynamo period
and parity of the two runs described
extensively in \cite{KKOBWKP16} and \cite{OlspertIEEE}.
For the cycle period estimation we used the $D^2$ statistic of \cite{Pelt83}, which
is extended to suit quasi-periodic time series. Additional to the frequency,
the statistic includes a free parameter called coherence time (or time-scale),
which quantifies the degree of non-periodicity.
$D^2$ spectrum for the azimuthal component of the magnetic field over the whole time interval
of the runs, depicted in \figu{fig:D2},
reveals that the basic cycle is indeed somewhat longer for Run~M2 than for M1.

In \cite{OlspertIEEE} we reported a peculiar feature of hemispheric asymmetry, namely the
cycle periods being different for different hemispheres, and this behaviour is now seen to
persist also with a different magnetic boundary condition.
The cycle periods for Run~M2 are 5.27~yr and 5.22~yr for north and
south, respectively. The corresponding values for Run~M1 are 5.17~yr
and 5.02~yr.
In the horizontal axis of the figure we also plot the ratio of the coherence time to the period $l_{\rm coh}$.
From this figure, it is evident that the cycle for Run~M2 is somewhat less coherent compared to that of M1.
The last thing to note from this figure is that the average cycle amplitude is slightly lower
for Run~M2 than for M1.

\begin{figure}[t!]
\begin{center}
\begin{minipage}{150mm}
\begin{center}
{\resizebox*{12.0cm}{!}{\includegraphics[trim=0.0cm 0.4cm 0.4cm 0.4cm,clip=true]{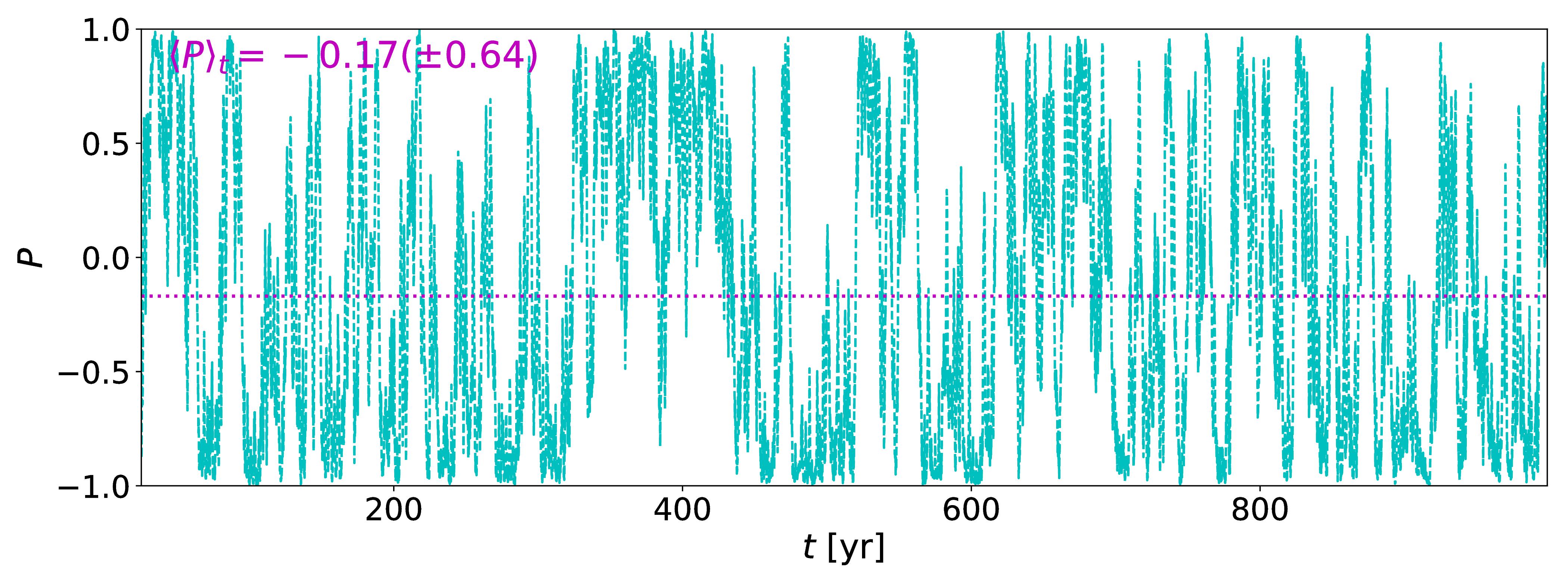}}}\\%
{\resizebox*{12.0cm}{!}{\includegraphics[trim=0.3cm 0.4cm 0.4cm 0.4cm,clip=true]{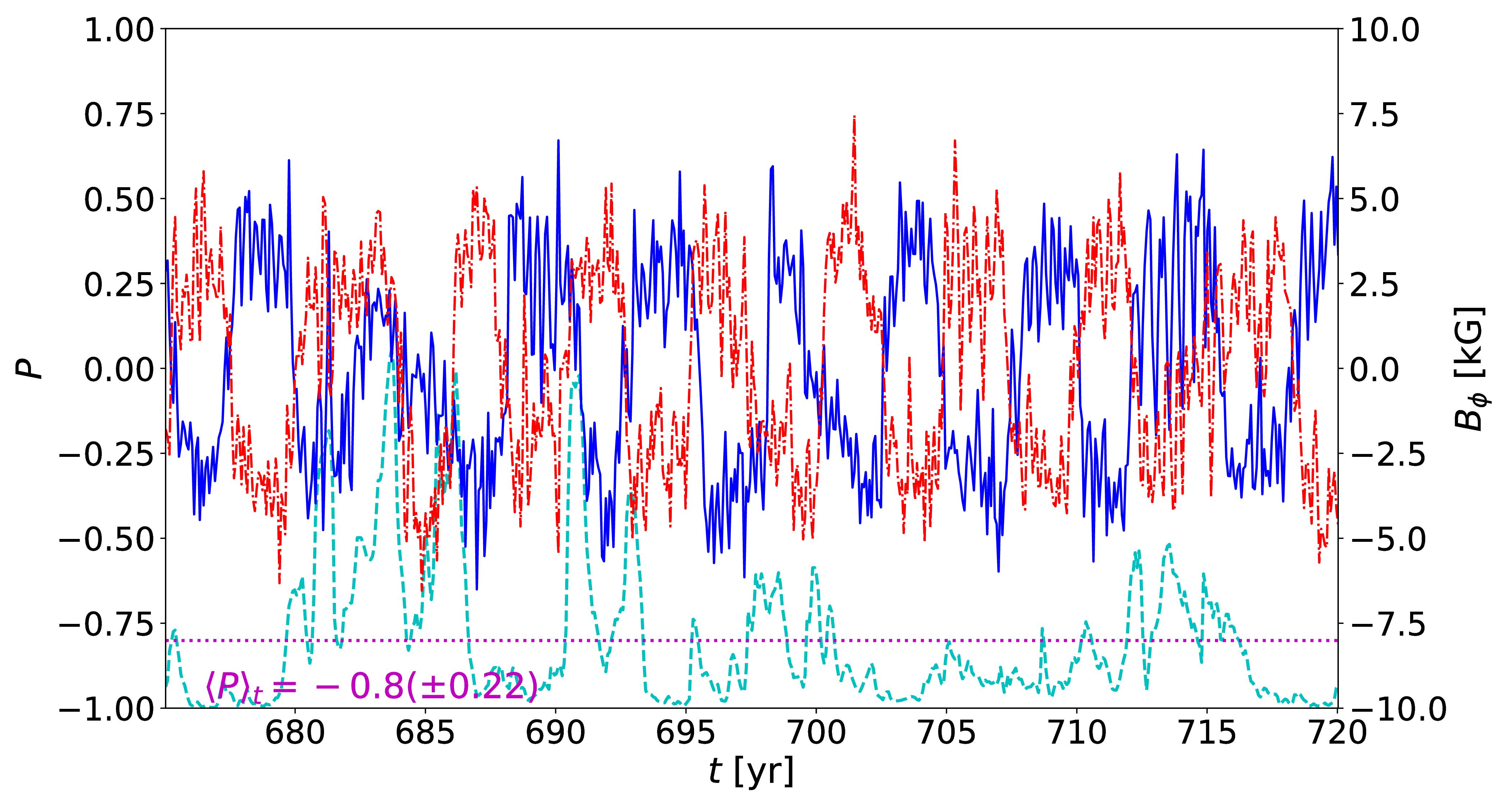}}}\\%
{\resizebox*{12.0cm}{!}{\includegraphics[trim=0.3cm 0.4cm 0.4cm 0.4cm,clip=true]{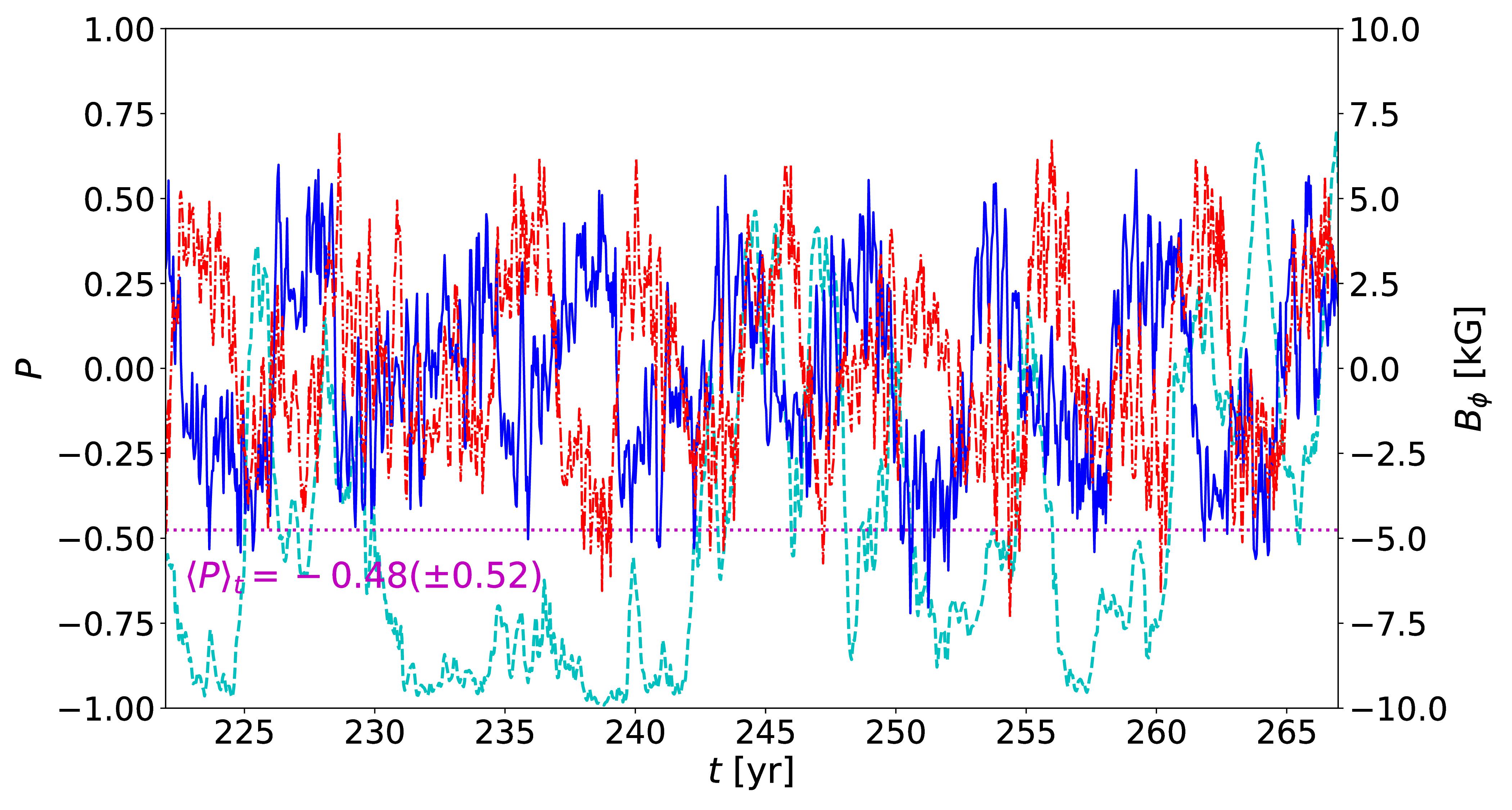}}}%
\caption{
Top panel: global instantaneous parity (cyan, dashed) and its temporal average
(magenta, dotted) from Run~M2.
Zoom-in over 45 years of same parity (cyan, dashed) and the 45 year temporal
average from Run~M1 (middle) and
similar period from Run~M2 (bottom), together with azimuthally
averaged toroidal magnetic field near the surface ($r=0.98R_\odot$) at
$\pm25\degree$ (blue, solid: north, red, dash-dotted: south) (colour online).
}%
\label{fig:parity}
\end{center}
\end{minipage}
\end{center}
\end{figure}

We have over 1000 years of data from Run~M1 and almost 1000 years for M2.
More detailed comparison of the full data sets including test-field
analysis is planned elsewhere.
In the top panel of \figu{fig:parity} we provide the time evolution of the
global parity for the full duration of Run~M2 for comparison with Fig~13(a) of
\cite{KKOBWKP16}, where the first 440 years of Run~M1 was presented.
Parity is a measure of the equatorial symmetry for the azimuthally
averaged magnetic field, defined as
\begin{equation}\label{eq:parity}
P\,=\,\frac{E_{\rm even}-E_{\rm odd}}{E_{\rm even}+E_{\rm odd}}\,,
\end{equation}
where $E_{\rm even}(E_{\rm odd})$ is the energy of the quadrupolar or symmetric
(dipolar or antisymmetric) mode of the magnetic field.
The temporal average of the global parity, which fluctuates between $\pm1$ is
$\langle P\rangle_t=-0.17$ with standard deviation $\sigma_P=0.64$ for Run~M2.
For Run~M1 up to about 440 years, \cite{KKOBWKP16} obtained
$\langle P\rangle_t=-0.15$ with standard deviation unreported, but it is
evident that the difference is not statistically significant.
If we define an error estimate as
\[
\epsilon\,=\,{\sigma_P}\Big/{\sqrt{N_{\rm cycle}}}\,,
\] then
we obtain $\epsilon=0.089$ and 0.053 for M1 and M2, respectively.

For direct comparison we have the lower two panels of \Figu{fig:parity} showing
the global parity during the 45 year solar-like intervals selected from both
Runs~M1 (middle) and M2 (lower), as well as the azimuthally averaged
toroidal field from latitudes $\pm 25\degree$ near the surface
at $r=0.98R_\odot$.
The time averaged parity during this brief interval is more strongly
dipolar with $\langle P\rangle_t=-0.8$ and $-0.48$, respectively, for Run~M1 and M2.

\section{Conclusions}

We have studied the influence of varying the imposed luminosity, changing the centrifugal force,
and adopting several thermal and magnetic boundary conditions on the solutions of HD
and MHD convection simulations in semi-global wedge geometry. We find
that changing the luminosity by an order of magnitude has a minor
influence on the large-scale quantities and that the fluctuations of
velocity and thermodynamic variables follow the expected power law
scalings \citep[e.g.][]{BCNS05}. Similarly, the centrifugal force has
only a minor influence
on the results, provided that its magnitude in comparison with the
acceleration due to gravity is still similar to that in real stars.
These results give us confidence that the fully compressible approach
taken with the {\sc Pencil Code} is indeed valid and offers certain
advantages, such as the inclusion of the not hopelessly disparate timescales
\citep[e.g.][]{KMCWB13}, over anelastic methods. However, a detailed
benchmark between anelastic and fully compressible codes would still
be desirable.

The most significant changes occur with the treatment of the
thermodynamics near the upper boundary. Cooling toward a fixed profile
of temperature near the surface
leads to a much more anisotropic convective heat flux than in cases
where an artificial radiative flux is extracted at the surface. These
results are insensitive to the thermal BC.
In the Sun the surface flux and temperature are almost independent of
latitude due to the vigorously mixed and rotationally weakly
affected surface layers. The current results suggest that until
simulations can capture the dynamics of these surface layers
self-consistently, great care has to be taken with the
parameterisation of the physics and the BCs that are imposed in the
current simulations.

The two adopted magnetic boundary conditions produce dynamo solutions that
are nearly identical. The only affected properties of the dynamo models are the cycle frequency and
the regularity of the basic dynamo mode. With the boundary condition that ensures vanishing
horizontal currents (vJ) at the bottom boundary, a somewhat longer solar-like cycle is produced,
while its coherence length (the time scale over which the cycle frequency remains stable),
measured by the $D^2$ statistics, is shorter than in the run with the vE boundary condition. The
cycle reported earlier by \cite{KKOBWKP16} from the {\sc Pencil Code} millennium simulation
was around 4.9 years, roughly five times too short in comparison to the Sun. Hence, even though
the new vJ boundary condition changes the cycle period into a more realistic direction, this
 change is far too subtle to bring the values into a realistic regime.

\section*{Acknowledgement}{
The anonymous referees are acknowledged for their constructive comments
on the paper.
The authors wish to acknowledge CSC -- IT Center for Science, who
are administered by the Finnish Ministry of Education; of Espoo,
Finland,
for computational resources. We also acknowledge the allocation of
computing resources through the Gauss Center for Supercomputing for
the Large-Scale computing project ``Cracking the Convective
Conundrum'' in the Leibniz Supercomputing Centre's SuperMUC
supercomputer in Garching, Germany. This work was supported in part by
the Deutsche Forschungsgemeinschaft Heisenberg programme (grant
No.\ KA 4825/1-1; PJK),
the Academy of Finland ReSoLVE Centre of Excellence (grant
No.\ 272157; MJK, PJK, FAG, NO), the NSF Astronomy and Astrophysics
Grants Program (grant 1615100), and the University of Colorado through
its support of the George Ellery Hale visiting faculty appointment.}

\bibliographystyle{gGAF}

\begin{thebibliography}{58}
\providecommand{\natexlab}[1]{#1}

\bibitem[\protect\citeauthoryear{{Augustson} {\itshape{et~al.}}}{2015}]{ABMT15}
{Augustson}, K., {Brun}, A.S., {Miesch}, M. and {Toomre}, J., {Grand minima and
  equatorward propagation in a cycling stellar convective dynamo}. {\itshape
  \apj}, 2015, \textbf{809}, 149.

\bibitem[\protect\citeauthoryear{{Augustson}
  {\itshape{et~al.}}}{2012}]{ABBMT12}
{Augustson}, K.C., {Brown}, B.P., {Brun}, A.S., {Miesch}, M.S. and {Toomre},
  J., {Convection and differential rotation in F-type stars}. {\itshape \apj},
  2012, \textbf{756}, 169.

\bibitem[\protect\citeauthoryear{{Barekat} and {Brandenburg}}{2014}]{BB14}
{Barekat}, A. and {Brandenburg}, A., {Near-polytropic stellar simulations with
  a radiative surface}. {\itshape \aap}, 2014, \textbf{571}, A68.

\bibitem[\protect\citeauthoryear{{Beaudoin}
  {\itshape{et~al.}}}{2018}]{2018ApJ...859...61B}
{Beaudoin}, P., {Strugarek}, A. and {Charbonneau}, P., {Differential rotation
  in solar-like convective envelopes: Influence of overshoot and magnetism}.
  {\itshape \apj}, 2018, \textbf{859}, 61.

\bibitem[\protect\citeauthoryear{{Brandenburg}}{2003}]{B03}
  {Brandenburg}, A., { Computational aspects of astrophysical MHD and
    turbulence}; in {\itshape {Advances in Nonlinear Dynamics}. Edited
    by Ferriz-Mas, A. and N{\'u}{\~n}ez, M.}, p. 269, 2003 (Taylor and
  Francis: London).

\bibitem[\protect\citeauthoryear{{Brandenburg}}{2016}]{Br16}
{Brandenburg}, A., {Stellar mixing length theory with entropy rain}. {\itshape
  \apj}, 2016, \textbf{832}, 6.

\bibitem[\protect\citeauthoryear{{Brandenburg}
  {\itshape{et~al.}}}{2005}]{BCNS05}
{Brandenburg}, A., {Chan}, K.L., {Nordlund}, {\AA}. and {Stein}, R.F., {Effect
  of the radiative background flux in convection}. {\itshape AN}, 2005,
  \textbf{326}, 681--692.

\bibitem[\protect\citeauthoryear{{Brandenburg} and {Dobler}}{2002}]{BD02}
{Brandenburg}, A. and {Dobler}, W., {Hydromagnetic turbulence in computer
  simulations}. {\itshape Comp. Phys. Comm.}, 2002, \textbf{147}, 471--475.

\bibitem[\protect\citeauthoryear{{Brandenburg}
  {\itshape{et~al.}}}{1992}]{BMT92}
{Brandenburg}, A., {Moss}, D. and {Tuominen}, I., {Stratification and
  thermodynamics in mean-field dynamos}. {\itshape \aap}, 1992, \textbf{265},
  328--344.

\bibitem[\protect\citeauthoryear{{Brandenburg}
  {\itshape{et~al.}}}{2000}]{2000gac..conf...85B}
{Brandenburg}, A., {Nordlund}, A. and {Stein}, R.F., {Astrophysical convection
  and dynamos}; in {\itshape Geophysical and astrophysical convection,
  contributions from a workshop sponsored by the Geophysical Turbulence Program
  at the National Center for Atmospheric Research, October, 1995. Edited by
  Peter A. Fox and Robert M. Kerr. Published by Gordon and Breach Science
  Publishers, The Netherlands, 2000, p. 85-105}, edited by P.A. {Fox} and R.M.
  {Kerr}, Aug., 2000, pp. 85--105.

\bibitem[\protect\citeauthoryear{{Brown} {\itshape{et~al.}}}{2008}]{BBBMT08}
{Brown}, B.P., {Browning}, M.K., {Brun}, A.S., {Miesch}, M.S. and {Toomre}, J.,
  {Rapidly rotating suns and active nests of convection}. {\itshape \apj},
  2008, \textbf{689}, 1354--1372.

\bibitem[\protect\citeauthoryear{{Brun} {\itshape{et~al.}}}{2004}]{BMT04}
{Brun}, A.S., {Miesch}, M.S. and {Toomre}, J., {Global-scale turbulent
  convection and magnetic dynamo action in the solar envelope}. {\itshape
  \apj}, 2004, \textbf{614}, 1073--1098.

\bibitem[\protect\citeauthoryear{{Brun} {\itshape{et~al.}}}{2011}]{BMT11}
{Brun}, A.S., {Miesch}, M.S. and {Toomre}, J., {Modeling the dynamical coupling
  of solar convection with the radiative interior}. {\itshape \apj}, 2011,
  \textbf{742}, 79.

\bibitem[\protect\citeauthoryear{{Deardorff}}{1966}]{De66}
{Deardorff}, J.W., {The counter-gradient heat flux in the lower atmosphere and
  in the laboratory.}. {\itshape J. Atmosph. Sci.}, 1966, \textbf{23},
  503--506.

\bibitem[\protect\citeauthoryear{{Fan} and {Fang}}{2014}]{FF14}
{Fan}, Y. and {Fang}, F., {A simulation of convective dynamo in the solar
  convective envelope: Maintenance of the solar-like differential rotation and
  emerging flux}. {\itshape \apj}, 2014, \textbf{789}, 35.

\bibitem[\protect\citeauthoryear{{Featherstone} and {Hindman}}{2016}]{FH16}
{Featherstone}, N.A. and {Hindman}, B.W., {The spectral amplitude of stellar
  convection and its scaling in the high-Rayleigh-number regime}. {\itshape
  \apj}, 2016, \textbf{818}, 32.

\bibitem[\protect\citeauthoryear{{Gastine} {\itshape{et~al.}}}{2012}]{GDW12}
{Gastine}, T., {Duarte}, L. and {Wicht}, J., {Dipolar versus multipolar
  dynamos: the influence of the background density stratification}. {\itshape
  \aap}, 2012, \textbf{546}, A19.

\bibitem[\protect\citeauthoryear{{Gastine} and
  {Wicht}}{2012}]{2012Icar..219..428G}
{Gastine}, T. and {Wicht}, J., {Effects of compressibility on driving zonal
  flow in gas giants}. {\itshape \icarus}, 2012, \textbf{219}, 428--442.

\bibitem[\protect\citeauthoryear{{Gastine} {\itshape{et~al.}}}{2014}]{GYMRW14}
{Gastine}, T., {Yadav}, R.K., {Morin}, J., {Reiners}, A. and {Wicht}, J., {From
  solar-like to antisolar differential rotation in cool stars}. {\itshape
  \mnras}, 2014, \textbf{438}, L76--L80.

\bibitem[\protect\citeauthoryear{{Gent} {\itshape{et~al.}}}{2017}]{GKW17}
{Gent}, F.A., {K{\"a}pyl{\"a}}, M.J. and {Warnecke}, J., {Long-term variations
  of turbulent transport coefficients in a solarlike convective dynamo
  simulation}. {\itshape Astronomische Nachrichten}, 2017, \textbf{338},
  885--895.

\bibitem[\protect\citeauthoryear{{Guerrero}
  {\itshape{et~al.}}}{2016}]{GSdGDPKM15}
{Guerrero}, G., {Smolarkiewicz}, P.K., {de Gouveia Dal Pino}, E.M.,
  {Kosovichev}, A.G. and {Mansour}, N.N., {On the role of tachoclines in solar
  and stellar dynamos}. {\itshape \apj}, 2016, \textbf{819}, 104.

\bibitem[\protect\citeauthoryear{{Hotta}}{2017}]{2017ApJ...843...52H}
{Hotta}, H., {Solar overshoot region and small-scale dynamo with realistic
  energy flux}. {\itshape \apj}, 2017, \textbf{843}, 52.

\bibitem[\protect\citeauthoryear{{Hotta} {\itshape{et~al.}}}{2014}]{HRY14}
{Hotta}, H., {Rempel}, M. and {Yokoyama}, T., {High-resolution calculations of
  the solar global convection with the reduced speed of sound technique. I. The
  structure of the convection and the magnetic field without the rotation}.
  {\itshape \apj}, 2014, \textbf{786}, 24.

\bibitem[\protect\citeauthoryear{{Hotta} {\itshape{et~al.}}}{2015}]{HRY15a}
{Hotta}, H., {Rempel}, M. and {Yokoyama}, T., {High-resolution calculation of
  the solar global convection with the reduced speed of sound technique. II.
  Near surface shear layer with the rotation}. {\itshape \apj}, 2015,
  \textbf{798}, 51.

\bibitem[\protect\citeauthoryear{{Hotta} {\itshape{et~al.}}}{2012}]{HRYIF12}
{Hotta}, H., {Rempel}, M., {Yokoyama}, T., {Iida}, Y. and {Fan}, Y., {Numerical
  calculation of convection with reduced speed of sound technique}. {\itshape
  \aap}, 2012, \textbf{539}, A30.

\bibitem[\protect\citeauthoryear{{Hurlburt} {\itshape{et~al.}}}{1984}]{HTM84}
{Hurlburt}, N.E., {Toomre}, J. and {Massaguer}, J.M., {Two-dimensional
  compressible convection extending over multiple scale heights}. {\itshape
  \apj}, 1984, \textbf{282}, 557--573.

\bibitem[\protect\citeauthoryear{{K{\"a}pyl{\"a}}
  {\itshape{et~al.}}}{2016}]{KKOBWKP16}
{K{\"a}pyl{\"a}}, M.J., {K{\"a}pyl{\"a}}, P.J., {Olspert}, N., {Brandenburg},
  A., {Warnecke}, J., {Karak}, B.B. and {Pelt}, J., {Multiple dynamo modes as a
  mechanism for long-term solar activity variations}. {\itshape \aap}, 2016,
  \textbf{589}, A56.

\bibitem[\protect\citeauthoryear{{K{\"a}pyl{\"a}}
  {\itshape{et~al.}}}{2014}]{KKB14}
{K{\"a}pyl{\"a}}, P.J., {K{\"a}pyl{\"a}}, M.J. and {Brandenburg}, A.,
  {Confirmation of bistable stellar differential rotation profiles}. {\itshape
  \aap}, 2014, \textbf{570}, A43.

\bibitem[\protect\citeauthoryear{{K{\"a}pyl{\"a}}
  {\itshape{et~al.}}}{2017{\natexlab{a}}}]{KKOWB16}
{K{\"a}pyl{\"a}}, P.J., {K{\"a}pyl{\"a}}, M.J., {Olspert}, N., {Warnecke}, J.
  and {Brandenburg}, A., {Convection-driven spherical shell dynamos at varying
  Prandtl numbers}. {\itshape \aap}, 2017{\natexlab{a}}, \textbf{599}, A5.

\bibitem[\protect\citeauthoryear{{K{\"a}pyl{\"a}}
  {\itshape{et~al.}}}{2010}]{KKBMT10}
{K{\"a}pyl{\"a}}, P.J., {Korpi}, M.J., {Brandenburg}, A., {Mitra}, D. and
  {Tavakol}, R., {Convective dynamos in spherical wedge geometry}. {\itshape
  Astron. Nachr.}, 2010, \textbf{331}, 73.

\bibitem[\protect\citeauthoryear{{K{\"a}pyl{\"a}}
  {\itshape{et~al.}}}{2011{\natexlab{a}}}]{KMB11}
{K{\"a}pyl{\"a}}, P.J., {Mantere}, M.J. and {Brandenburg}, A., {Effects of
  stratification in spherical shell convection}. {\itshape Astron. Nachr.},
  2011{\natexlab{a}}, \textbf{332}, 883.

\bibitem[\protect\citeauthoryear{{K{\"a}pyl{\"a}}
  {\itshape{et~al.}}}{2012}]{KMB12}
{K{\"a}pyl{\"a}}, P.J., {Mantere}, M.J. and {Brandenburg}, A., {Cyclic magnetic
  activity due to turbulent convection in spherical wedge geometry}. {\itshape
  \apjl}, 2012, \textbf{755}, L22.

\bibitem[\protect\citeauthoryear{{K{\"a}pyl{\"a}}
  {\itshape{et~al.}}}{2013}]{KMCWB13}
{K{\"a}pyl{\"a}}, P.J., {Mantere}, M.J., {Cole}, E., {Warnecke}, J. and
  {Brandenburg}, A., {Effects of enhanced stratification on equatorward dynamo
  wave propagation}. {\itshape \apj}, 2013, \textbf{778}, 41.

\bibitem[\protect\citeauthoryear{{K{\"a}pyl{\"a}}
  {\itshape{et~al.}}}{2011{\natexlab{b}}}]{KMGBC11}
{K{\"a}pyl{\"a}}, P.J., {Mantere}, M.J., {Guerrero}, G., {Brandenburg}, A. and
  {Chatterjee}, P., {Reynolds stress and heat flux in spherical shell
  convection}. {\itshape \aap}, 2011{\natexlab{b}}, \textbf{531}, A162.

\bibitem[\protect\citeauthoryear{{K{\"a}pyl{\"a}}
  {\itshape{et~al.}}}{2017{\natexlab{b}}}]{2017ApJ...845L..23K}
{K{\"a}pyl{\"a}}, P.J., {Rheinhardt}, M., {Brandenburg}, A., {Arlt}, R.,
  {K{\"a}pyl{\"a}}, M.J., {Lagg}, A., {Olspert}, N. and {Warnecke}, J.,
  {Extended subadiabatic layer in simulations of overshooting convection}.
  {\itshape \apjl}, 2017{\natexlab{b}}, \textbf{845}, L23.

\bibitem[\protect\citeauthoryear{{K{\"a}pyl{\"a}}
  {\itshape{et~al.}}}{2019}]{2018arXiv180305898K}
{K{\"a}pyl{\"a}}, P.J., {Viviani}, M., {K{\"a}pyl{\"a}}, M.J. and
  {Brandenburg}, A., {Effects of a subadiabatic layer on convection and dynamos
  in spherical wedge simulations}. {\itshape arXiv:1803.05898}, 2019.

\bibitem[\protect\citeauthoryear{{Kitchatinov}
  {\itshape{et~al.}}}{1994}]{KPR94}
{Kitchatinov}, L.L., {Pipin}, V.V. and {R\"udiger}, G., {Turbulent viscosity,
  magnetic diffusivity, and heat conductivity under the influence of rotation
  and magnetic field}. {\itshape Astron. Nachr.}, 1994, \textbf{315}, 157--170.

\bibitem[\protect\citeauthoryear{Krause and R{\"a}dler}{1980}]{KR80}
Krause, F. and R{\"a}dler, K.H., {\itshape {Mean-field magnetohydrodynamics and
  dynamo theory}},  1980 (Oxford: Pergamon Press).

\bibitem[\protect\citeauthoryear{{Kupka} and
  {Muthsam}}{2017}]{2017LRCA....3....1K}
{Kupka}, F. and {Muthsam}, H.J., {Modelling of stellar convection}. {\itshape
  Liv. Rev. Comp. Astrophys.}, 2017, \textbf{3}, 1.

\bibitem[\protect\citeauthoryear{{Mabuchi} {\itshape{et~al.}}}{2015}]{MMK15}
{Mabuchi}, J., {Masada}, Y. and {Kageyama}, A., {Differential rotation in
  magnetized and non-magnetized stars}. {\itshape \apj}, 2015, \textbf{806},
  10.

\bibitem[\protect\citeauthoryear{{Masada} {\itshape{et~al.}}}{2013}]{MK13}
{Masada}, Y., {Yamada}, K. and {Kageyama}, A., {Effects of penetrative
  convection on solar dynamo}. {\itshape \apj}, 2013, \textbf{778}, 11.

\bibitem[\protect\citeauthoryear{{Mitra} {\itshape{et~al.}}}{2009}]{MTBM09}
{Mitra}, D., {Tavakol}, R., {Brandenburg}, A. and {Moss}, D., {Turbulent
  dynamos in spherical shell segments of varying geometrical extent}. {\itshape
  \apj}, 2009, \textbf{697}, 923--933.

\bibitem[\protect\citeauthoryear{{Moffatt}}{1978}]{M78}
{Moffatt}, H.K., {\itshape {Magnetic field generation in electrically
  conducting fluids}},  1978 (Cambridge: Cambridge University Press).

\bibitem[\protect\citeauthoryear{{Nelson}
  {\itshape{et~al.}}}{2018}]{2018ApJ...859..117N}
{Nelson}, N.J., {Featherstone}, N.A., {Miesch}, M.S. and {Toomre}, J., {Driving
  solar giant cells through the self-organization of near-surface plumes}.
  {\itshape \apj}, 2018, \textbf{859}, 117.

\bibitem[\protect\citeauthoryear{{Olspert}
  {\itshape{et~al.}}}{2016}]{OlspertIEEE}
{Olspert}, N., {K{\"a}pyl{\"a}}, M.J. and {Pelt}, J., Method for estimating
  cycle lengths from multidimensional time series: Test cases and application
  to a massive "in silico" dataset; in {\itshape 2016 {IEEE} International
  Conference on Big Data, BigData 2016, Washington DC, USA, December 5-8,
  2016}, 2016, pp. 3214--3223.

\bibitem[\protect\citeauthoryear{{Pelt}}{1983}]{Pelt83}
{Pelt}, J., {Phase dispersion minimization methods for estimation of periods
  from unequally spaced sequences of data}; in {\itshape Statistical Methods in
  Astronomy}, edited by E.J. {Rolfe}, Vol.  201 of {\itshape ESA Special
  Publication}, Nov., 1983, pp. 37--42.

\bibitem[\protect\citeauthoryear{{Rempel}}{2005}]{Re05}
{Rempel}, M., {Solar differential rotation and meridional flow: The role of a
  subadiabatic tachocline for the Taylor-Proudman balance}. {\itshape \apj},
  2005, \textbf{622}, 1320--1332.

\bibitem[\protect\citeauthoryear{{R\"udiger}}{1989}]{R89}
{R\"udiger}, G., {\itshape {Differential rotation and stellar convection. Sun
  and solar-type stars}},  1989 (Berlin: Akademie Verlag).

\bibitem[\protect\citeauthoryear{{Schrinner}
  {\itshape{et~al.}}}{2005}]{SRSRC05}
{Schrinner}, M., {R{\"a}dler}, K.H., {Schmitt}, D., {Rheinhardt}, M. and
  {Christensen}, U., {Mean-field view on rotating magnetoconvection and a
  geodynamo model}. {\itshape Astron. Nachr.}, 2005, \textbf{326}, 245--249.

\bibitem[\protect\citeauthoryear{{Schrinner}
  {\itshape{et~al.}}}{2007}]{SRSRC07}
{Schrinner}, M., {R{\"a}dler}, K.H., {Schmitt}, D., {Rheinhardt}, M. and
  {Christensen}, U.R., {Mean-field concept and direct numerical simulations of
  rotating magnetoconvection and the geodynamo}. {\itshape Geophys. Astrophys.
  Fluid Dynam.}, 2007, \textbf{101}, 81--116.

\bibitem[\protect\citeauthoryear{{Simitev}
  {\itshape{et~al.}}}{2015}]{2015ApJ...810...80S}
{Simitev}, R.D., {Kosovichev}, A.G. and {Busse}, F.H., {Dynamo effects near the
  transition from solar to anti-solar differential rotation}. {\itshape \apj},
  2015, \textbf{810}, 80.

\bibitem[\protect\citeauthoryear{{Singh}
  {\itshape{et~al.}}}{1998}]{1998A&A...340..178S}
{Singh}, H.P., {Roxburgh}, I.W. and {Chan}, K.L., {A study of penetration at
  the bottom of a stellar convective envelope and its scaling relationships}.
  {\itshape \aap}, 1998, \textbf{340}, 178--182.

\bibitem[\protect\citeauthoryear{{Smolarkiewicz} and
  {Charbonneau}}{2013}]{SC13}
{Smolarkiewicz}, P.K. and {Charbonneau}, P., {EULAG, a computational model for
  multiscale flows: An MHD extension}. {\itshape J. Comp. Phys.}, 2013,
  \textbf{236}, 608--623.

\bibitem[\protect\citeauthoryear{{Tian}
  {\itshape{et~al.}}}{2009}]{2009MNRAS.398.1011T}
{Tian}, C.L., {Deng}, L.C. and {Chan}, K.L., {Numerical simulations of downward
  convective overshooting in giants}. {\itshape \mnras}, 2009, \textbf{398},
  1011--1022.

\bibitem[\protect\citeauthoryear{{Tremblay}
  {\itshape{et~al.}}}{2015}]{2015ApJ...799..142T}
{Tremblay}, P.E., {Ludwig}, H.G., {Freytag}, B., {Fontaine}, G., {Steffen}, M.
  and {Brassard}, P., {Calibration of the mixing-length theory for convective
  white dwarf envelopes}. {\itshape \apj}, 2015, \textbf{799}, 142.

\bibitem[\protect\citeauthoryear{{Warnecke} {\itshape{et~al.}}}{2014}]{WKKB14}
{Warnecke}, J., {K{\"a}pyl{\"a}}, P.J., {K{\"a}pyl{\"a}}, M.J. and
  {Brandenburg}, A., {On the cause of solar-like equatorward migration in
  global convective dynamo simulations}. {\itshape \apjl}, 2014, \textbf{796},
  L12.

\bibitem[\protect\citeauthoryear{{Warnecke} {\itshape{et~al.}}}{2016}]{WKKB16}
{Warnecke}, J., {K{\"a}pyl{\"a}}, P.J., {K{\"a}pyl{\"a}}, M.J. and
  {Brandenburg}, A., {Influence of a coronal envelope as a free boundary to
  global convective dynamo simulations}. {\itshape \aap}, 2016, \textbf{596},
  A115.

\bibitem[\protect\citeauthoryear{{Warnecke} {\itshape{et~al.}}}{2013}]{WKMB13}
{Warnecke}, J., {K{\"a}pyl{\"a}}, P.J., {Mantere}, M.J. and {Brandenburg}, A.,
  {Spoke-like differential rotation in a convective dynamo with a coronal
  envelope}. {\itshape \apj}, 2013, \textbf{778}, 141.

\bibitem[\protect\citeauthoryear{{Warnecke}
  {\itshape{et~al.}}}{2018}]{2018A&A...609A..51W}
{Warnecke}, J., {Rheinhardt}, M., {Tuomisto}, S., {K{\"a}pyl{\"a}}, P.J.,
  {K{\"a}pyl{\"a}}, M.J. and {Brandenburg}, A., {Turbulent transport
  coefficients in spherical wedge dynamo simulations of solar-like stars}.
  {\itshape \aap}, 2018, \textbf{609}, A51.

\bibitem[\protect\citeauthoryear{{Weiss} {\itshape{et~al.}}}{2004}]{WHTR04}
{Weiss}, A., {Hillebrandt}, W., {Thomas}, H.C. and {Ritter}, H., {\itshape {Cox
  and Giuli's principles of stellar structure}},  2004 (Cambridge, UK:
  Cambridge Scientific Publishers Ltd).
\end{thebibliography}
\markboth{\rm{P.J.\ K\"APYL\"A et al.}}{\rm GEOPHYSICAL AND ASTROPHYSICAL FLUID DYNAMICS}

\vspace{36pt}
\markboth{\rm{P.J.\ K\"APYL\"A et al.}}{\rm GEOPHYSICAL AND ASTROPHYSICAL FLUID DYNAMICS}
\appendices

\section{Units and conversion factors to physical units}
\label{app:units}

\noindent
The unit of time is given by the rotation period of the star:
\begin{eqnarray}
[t]\, = \,{2\pi}/{\Omega}\,,\label{equ:time}
\end{eqnarray}
where $\Omega$ is the angular velocity of the star. The unit of length
is given by the radius of the star:
\begin{eqnarray}
[x] \,=\, R\,.
\end{eqnarray}
The density is given in units of its initial value at the base of the
convection zone:
\begin{eqnarray}
[\rho]\, = \,\rho_{\rm bot}(t=0)\,.
\end{eqnarray}
The unit of velocity is constructed using $[t]$ and $[x]$:
\begin{eqnarray}
[U] \,= \,{[x]}\big/{[t]}\, = \,{\Omega R}\big/{2\pi}\,.
\end{eqnarray}
The unit of magnetic field is obtained from the definition of the
equipartition field strength:
\begin{eqnarray}
{B_{\rm eq}^2}\big/{\mu_0} \,=\, \rho {\bm U}^2 \hskip 12mm  \Longrightarrow \hskip 12mm B_{\rm eq}\, = \,\sqrt{\mu_0 \rho {\bm U}^2}\,.
\end{eqnarray}
Thus,
\begin{eqnarray}
[B]\, =\, \sqrt{\mu_0 [\rho] [U]^2}\,.
\end{eqnarray}
Let us consider a simulation targeted toward a star with a
particular luminosity and rotation rate. Then we assume that the
dimensionless time, velocity, density,
and magnetic fields are the same in the simulation as in the
target star. For
example, for time this means that:
\begin{align}
  t^{\rm sim}/[t]\, = \,t/[t]\hskip 12mm &\Longleftrightarrow \hskip 12mm t^{\rm sim}\Omsi/2\pi\,=\, t \Omega/2\pi \nonumber\\
  &\Longleftrightarrow \hskip 12mm  t\, =\, \frac{\Omsi}{\Omega} t^{\rm sim} \,\equiv\, c_t t^{\rm sim}\,,
\end{align}
which gives time in physical units with $c_t$ being the conversion
factor. The superscript `sim' refers to the quantities in code units
while quantities without superscripts refer to values in
physical units. Note that
$\Omsi$ is the rotation rate of the target star in code units.

Performing the same exercise for the density, velocity, and
magnetic fields yields
\begin{gather}
\rho \,=\, \frac{\rhobot}{\rhosib} \rhosi\,,  \hskip 25mm
U \,=\, \left(\frac{\Omega R}{\Omsi \Rsi} \right) \usi\,,\nonumber\\
B = \,\left[\frac{\mu_0\rhobot (\Omega R)^2}{\mu_0^{\rm sim} \rhosib (\Omsi \Rsi)^2} \right]^{1/2} \Bsi\,,
\end{gather}
where $\rhobot$ is the density at the bottom of the CZ in the
star in physical units.
Here $\rhosib$ and $\Rsi$ are the solar density at the base of the
convection zone and the solar radius in code units. Furthermore,
$\mu_0^{\rm sim}$ is the magnetic
permeability in code units. Thus the conversion factors are
\begin{align}
c_t \,=\,& \,\frac{\Omsi}{\Omega}\,,  &
c_\rho \,=\,& \,\frac{\rhobot}{\rhosib}\,,\nonumber\\
\qquad c_U\, =\,&\, \left(\frac{\Omega R}{\Omsi \Rsi} \right),&
c_B\, =\,&\, \left[\frac{\mu_0\rhobot (\Omega R)^2}{\mu_0^{\rm sim} \rhosib (\Omsi \Rsi)^2} \right]^{1/2}.\qquad
\end{align}
The conversion factors are then fully determined once $\Omsi$,
$\rhosib$, $\Rsi$, and $\mu_0^{\rm sim}$ are chosen. Typically the
last three are set to unity in code units:
\begin{eqnarray}
\rhosib\, =\, \Rsi\, = \,\mu_0^{\rm sim} \,= \,1\,,
\end{eqnarray}
whereas the value of $\Omsi$ depends on the rotation rate of the
target star and the factor by which the luminosity is enhanced.

\section*{Enhanced luminosity and scaling to stellar-equivalent rotational state}

\noindent
The dimensionless luminosity is given by
\begin{eqnarray}
\mathcal{L}\, = \,\frac{L}{\rho_{\rm bot} (GM)^{3/2} R^{1/2}}\,,
\end{eqnarray}
where $L$, $\rho_{\rm bot}$, $G$, $M$, and $R$ are the luminosity,
density at the bottom of the convection zone, gravitational constant,
mass and radius of the star, respectively.
In the code $GM$ is given by the input parameter \texttt{gravx} and
the luminosity is computed from the given flux $F_{\rm bot}$ at the
bottom boundary:
\begin{eqnarray}
L \,=\, 4\pi r_0^2 F_{\rm bot}\,,
\end{eqnarray}
where $r_0$ is the inner radius. Given that the fully compressible
formulation does not allow a realistic flux due to the short time
steps from sound waves, we typically use a much higher luminosity than
that of stars such as the Sun. The ratio of the luminosities
of the simulation and the target star is denoted as:
\begin{eqnarray}
\Lratio\, =\, \mathcal{L}_{\rm sim}/\mathcal{L}.
\end{eqnarray}
The convective velocity scales with the luminosity as $u\propto
L^{1/3}$; see \figu{fig:pMach}(a). This means that in order to
capture the same rotational
influence on the flow as in the Sun, the rotation rate must be
enhanced by the same factor as the velocities are amplified. We call
the resulting setup the stellar-equivalent rotational state and
correspondingly refer to the resulting value of $\Omega$ as the
stellar-equivalent value $\Omsi$.
Another time unit is need to represent $\Omsi$ in dimensionless form.
We use the acceleration due to gravity at the surface of the
star to construct this:
\begin{equation}
g\, = \frac{GM}{R^2}\ = \frac{[x]}{[t_{\rm alt}]^2} \hskip 12mm \Longrightarrow\hskip 12mm [t_{\rm alt}]\, = \,\left(\frac{R}{g}\right)^{\!1/2}\,,
\end{equation}
where $t_{\rm alt}$ is an alternative time unit, and
$[x]=R$ has been used.
Using $[\Omega]=2\pi/[t_{\rm alt}]$ and taking into account the
enhanced luminosity in the rotation rate in the simulations, we
obtain
\begin{gather}
\Omsi \left(\frac{\Rsi}{\gsi}\right)^{\!1/2} \,= \,\Lratio^{1/3} \Omega \left(\frac{R}{g}\right)^{\!1/2} \hskip 70mm\nonumber \\
\hskip 30mm  \Longleftrightarrow\hskip 12mm \Omsi \,= \,\Lratio^{1/3} \left(\frac{\gsi}{g}\frac{R}{\Rsi}\right)^{\!1/2} \Omega \,,
\end{gather}
with
\begin{eqnarray}
c_\Omega \,=\, \Lratio^{1/3} \left(\frac{\gsi}{g}\frac{R}{\Rsi}\right)^{1/2},
\end{eqnarray}
completing the conversion factors between physical and simulation
units. In the current study we use $\texttt{gravx} = \gsi=3$ in code units.

This setup can be understood literally as described above as a
solar-like star where the luminosity is greatly enhanced and where the
convective velocities are $\Lratio^{1/3}$ higher than in the Sun. On
the other hand, one can also interpret it as a star with a
sound speed (temperature) that is $\Lratio^{1/3}$ ($\Lratio^{2/3}$)
lower than in the Sun. Neither case corresponds to a real star, but
the current setup offers clear numerical advantages. With a Mach
number on the order of $10^{-2}\ldots0.1$, the acoustic and convective
time scales are not too far apart for the former to become dominant in
the time step calculation. The higher luminosity also allows runs that
can be thermally relaxed which cannot be performed with a realistic
luminosity.

\end{document}